\documentclass[a4paper,10pt,sort,compress]{elsarticle}

\usepackage[UTF8]{ctex}
\usepackage{graphicx}
\usepackage{amsfonts}
\usepackage{textcomp}
\usepackage{amssymb}
\usepackage{mathrsfs}
\usepackage{amsmath}
\usepackage{xcolor}
\usepackage{float}
\usepackage{empheq}
\usepackage[english]{babel} 
\usepackage[left=2.0cm,right=2.0cm,top=2.5cm,bottom=2.5cm]{geometry}

\usepackage{setspace}
\usepackage[switch]{lineno} %
\modulolinenumbers[1]       %
\newcommand{\convo}{\ast}

\journal{IJMS}

\begin{document}

\setstretch{1.5}
\onehalfspacing

\begin{frontmatter}

\title{Wrinkling of Randomly Heterogeneous Film--Substrate Systems}

\author[pku_me]{Xinyu Xing\fnref{equal}}
\author[pku_eng]{Liyu Zhong\fnref{equal}}
\author[xjtu]{Feng Deng\corref{cor1}}
\ead{dengfchn@xjtu.edu.cn}
\author[pku_me,pku_eng]{Sheng Mao\corref{cor1}}
\ead{maosheng@pku.edu.cn}
\fntext[equal]{These authors contributed equally to this work.}
\cortext[cor1]{Corresponding author.}

\address[pku_me]{School of Mechanics and Engineering Science, Peking University, Beijing 100871, P. R. China}
\address[pku_eng]{Department of Mechanics and Engineering Science, College of Engineering, Peking University, Beijing 100871, P. R. China}
\address[xjtu]{School of Aerospace Engineering, Xi'an Jiaotong University, Xi'an 710049, P. R. China}

\begin{abstract}

Wrinkling instabilities in stiff films bonded to compliant substrates are highly sensitive to spatial variations in film stiffness, which commonly arise from thickness fluctuations, compositional heterogeneity, microstructural defects, and processing-induced non-uniformity.
This paper develops a homogenized instability framework for one-dimensional film--substrate systems with random bending stiffness.
The heterogeneous stability problem is reformulated as a Lippmann--Schwinger equation for the curvature field, and a strong-contrast expansion (SCE) is constructed through a local--nonlocal kernel decomposition and a cavity-field representation.
Truncation up to third order yields an effective polarizability and a Dyson-type effective dispersion relation, from which the critical load and critical wavenumber are predicted. 
The stiffness field is modeled as an exponentially mapped Gaussian random field, for which the required two- and three-point connected statistics are obtained in closed form.
Theoretical predictions are validated against generalized eigenvalue calculations and Fourier spectral simulations.
The results show that increasing stiffness contrast lowers the critical load and shifts the unstable mode toward higher wavenumbers, leading to shorter wrinkling wavelengths. 
In the weak-contrast regime, the instability threshold follows a universal quadratic scaling $N_c^{(0)}-N_c\sim\varepsilon^2$, while in the moderate- and strong-contrast regimes, the third-order SCE gives more robust predictions than the second-order approximation. 
The wavelength-selection mechanism is governed by the ratio between the dominant material wavelength $\lambda^\ast$ and the harmonic-mean reference wrinkling wavelength $\lambda_H$. 
When $\lambda^\ast/\lambda_H<1$, the actual wrinkling wavelength is well predicted by the harmonic-mean model.
Overall, the proposed framework provides a mechanics-based basis for reliability assessment and structural design in statistically heterogeneous film--substrate systems.
\end{abstract}


\end{frontmatter}

\section{Introduction}
Film--substrate systems subjected to compression commonly exhibit wrinkling instabilities, producing periodic surface morphologies with characteristic wavelengths and amplitudes~\cite{chen2004herringbone,BlosseyNatureMaterials2003selfcleaning, CrosbyAM2006fabricating, BrauSoftMatter2013Wrinkle,zhao2020multilayered, DengScienceAdvance2025Nonliving}. 
The compressive stress may be generated by a mismatch in thermal expansion~\cite{bowden1998spontaneous, songJAP2008analytical, barnes2023surface},  swelling~\cite{kimAM2011solventresponsive, vandeparrePRL2008wrinkling, velankarACS2012swellinginduced}, or residual stresses introduced during fabrication~\cite{pundtActa2004adhesion,mukherjee2025deformation}.
Wrinkling is widely observed in natural morphogenesis, including surface patterns in fruits and vegetables~\cite{yin2008stressdriven,yin2009anisotropic}, wrinkles in skin~\cite{cerda2003geometry,dagdeviren2015conformal,chavoshnejad2021effect}, and geological folding~\cite{hudleston2010information}.
It has also been exploited in engineering applications such as tunable optics~\cite{leeAM2010switchable, wang2011unlocking,jung2011dynamically,shen2014random, wu2020harnessing,chen2021flower}, stretchable and flexible electronic devices~\cite{khangScience2006stretchable, xuAM2014mechanically, wangAPL2009wrinkled, wu2019stretchable,  xieAFM2021delamination,chen2025effect}, and biomedical systems~\cite{wuJMPS2026instability, LumelskyIEEE2001Sensitiveskin, nathan2000amorphous}.
These applications require a quantitative understanding of the onset of instability, wavelength selection, mode competition, and post-buckling evolution.

The mechanics of spatially homogeneous film--substrate systems is now relatively well established. 
Early analyses identified the competition between film bending and substrate resistance as the basic mechanism governing the instability threshold~\cite{biot1937bending}. 
Subsequent stability analyses related the critical load and wavelength to film thickness, elastic mismatch, substrate thickness, interfacial conditions, and loading state~\cite{chen2004herringbone,CrosbyAM2006fabricating}.
Nonlinear energy minimization and asymptotic analyses further predicted wrinkle amplitudes and the competition among post-buckling modes~\cite{audolyJMPS2008buckling1,audolyJMPS2008buckling2,audolyJMPS2008buckling3,cai2011periodic,kordolemis2025wrinkling,ren2026novel}. 
At larger compressive strains, wrinkles may undergo period doubling~\cite{cao2015towards,saha2017geometric,yu2010tunable}, folding~\cite{BrauSoftMatter2013Wrinkle,zhao2017multimodal,nikraveshThinWall2024wrinkle,nagashima2026elastocapillary} and ridge~\cite{zang2012localized,jin2015mechanics,guan2022compressioninduced}.
Recent studies have extended the classical framework to dynamic loading~\cite{box2019dynamics,abighanem2019wrinkles,macnider2023dynamic,zhouJAM2024dynamic,simoes2026dynamics}, substrate viscoelasticity~\cite{yu2017wrinkling,liu2020radial,liu2022radial}, prestretch~\cite{AugusteSoftmatter2014role,cao2015towards,qiu2025stretchinduced} and multilayered architectures~\cite{bakiler2022wrinkling,wang2025roles,kwak2026wrinkling}.
Spectral methods~\cite{huang2005nonlinear,yin2018surface}, nonlinear finite-element simulations~\cite{shao2016curvature,miyoshiJMPS2021bifurcation,kikuchiJMPS2022diversity,nikraveshThinWall2024wrinkle,SamyJMPS2025electromechanical} and other numerical methods~\cite{shen2024wrinkling,yang2025programmable, XuJMPS2025nonlinear,guo2026path} have also been developed to resolve large-domain pattern evolution and multiple post-buckling branches. 

Practical film and substrate are rarely perfectly homogeneous.
Most existing studies on heterogeneous systems focus on deterministic heterogeneity, including periodic or graded stiffness modulation~\cite{yang2014analysis, wangAPL2016wrinkling, wang2023firstorder, xueIJSS2020articulated, fuIJNM2024wrinkling, shen2025wrinkling,zhao2026nonlinear}, thickness gradients~\cite{croll2012pattern, li2013patterned, yuACS2015tunable, wangJAM2015surface,yu2019controlled,yu2022tailoring,wang2025peeling} and regularly arranged microstructures~\cite{efimenko2005nested, lee2012anisotropic, huSoftMatter2016wrinkling, nagashima2023bioinspired,sun2024wrinkling,zhang2019buckling}. 
In these systems, the spatial distribution of material properties is prescribed in advance, allowing the wrinkle morphology to be controlled through the location, geometry, or periodicity of the heterogeneity~\cite{zhou2026optimizing}. 
Manufacturing-induced variations, however, are generally irregular and cannot be represented by a single deterministic profile~\cite{baizhikovaJAM2022stochastic, liuJMPS2024surface}.
Instead, spatial fluctuations in stiffness, thickness, or residual stress are more naturally modeled as random fields.

Stochastic homogenization and effective-medium theories provide a natural starting point for describing the macroscopic response of such random materials \cite{li2019predicting,roy2022analytical,dhar2025effective,xue2026hyperelastic,zhang2026secondorder,shi2026representative,nguyen2026symmetries}.
Classical variational bounds \cite{hashin1963variational,beran1965use,milton1981bounds} and strong-contrast expansions~\cite{torquato1985effective,Torquato2003strongcontrast,Torquato2018designing,torquato2025accurate} relate effective material properties to statistical descriptors of the underlying microstructure, particularly two- and higher-order correlation functions. These approaches have been applied extensively to static effective properties, including conductivity, elasticity, and transport. 
More recently, for heterogeneous materials represented by continuous random fields, Zhong et al.~\cite{zhong2025contrast} reformulated variable-coefficient governing equations as a Lippmann--Schwinger integral equation  and introduced a cavity-field renormalization to improve the convergence of the associated Neumann expansion.
For Gaussian random fields, the resulting corrections can be expressed in terms of low-order statistical descriptors, such as the covariance function.
Wrinkling, however, differs fundamentally from static homogenization problems. 
It is a spectral instability problem in which the critical threshold is determined by the minimum of an effective dispersion relation. 
Random stiffness fluctuations introduce wavenumber coupling and modify not only the instability threshold but also the critical mode structure. 
Therefore, a direct application of standard homogenization formulas is not sufficient.

To address this issue, this work develops a strong-contrast homogenization framework for one-dimensional film--substrate systems with randomly heterogeneous bending stiffness.
The substrate response is represented by a unified Fourier-domain kernel, which covers Winkler/liquid foundations and, after condensation of tangential--normal interfacial coupling, elastic half-space substrates.
The heterogeneous stability problem is reformulated as a Lippmann--Schwinger equation for the curvature field. By introducing a local--nonlocal kernel decomposition and a cavity-field representation, we derive a strong-contrast expansion for the effective polarizability and obtain a Dyson-type effective dispersion relation for the critical load and critical wavenumber. 
For exponentially mapped Gaussian stiffness fields, the required two- and three-point connected statistics are derived in closed form. 
The theoretical predictions are validated against generalized eigenvalue calculations and Fourier spectral simulations, and the effects of stiffness contrast and material correlation length on the instability threshold, wavelength selection, and mode localization are systematically examined.
The framework provides a predictive tool for assessing instability reliability and guiding structural design in heterogeneous thin-film systems.

The rest of this paper is organized to progressively connect the mechanics of heterogeneous wrinkling with a statistically informed homogenized instability criterion. 
Section~2 formulates the one-dimensional film--substrate model, introduces the unified substrate-kernel representation, and summarizes the numerical eigenvalue and spectral methods used for validation.  
Section~3 develops the Lippmann--Schwinger formulation, the strong-contrast expansion, and the homogenized instability criterion.
Section~4 specifies the exponentially mapped Gaussian random stiffness model and derives the closed-form two- and three-point connected statistics required by the SCE.  
Section~5 compares the theory with numerical simulations and discusses how stiffness contrast and material correlation length affect the critical load, wavelength selection, spectral complexity, and spatial localization of the critical mode. 
Finally, the conclusions summarize the principal findings, clarify the physical implications of the proposed framework, and outline possible extensions.

\section{Theoretical Model and Governing Equations}

This section establishes the mechanical model used throughout the paper. 
We first define the one-dimensional film--substrate configuration, the random bending-stiffness field, and the basic assumptions adopted in the thin-film description. 
We then derive the governing equations for both Winkler-type and elastic half-space substrates and introduce a unified scalar substrate kernel that allows these cases to be treated within the same instability framework. 
Finally, the homogeneous benchmark and the numerical methods for heterogeneous systems are summarized, providing reference solutions for validating the strong-contrast theory developed in the following sections.

\subsection{Problem Description and Basic Assumptions}

\begin{figure}[ht]
    \centering
    \includegraphics[width=0.9\textwidth]{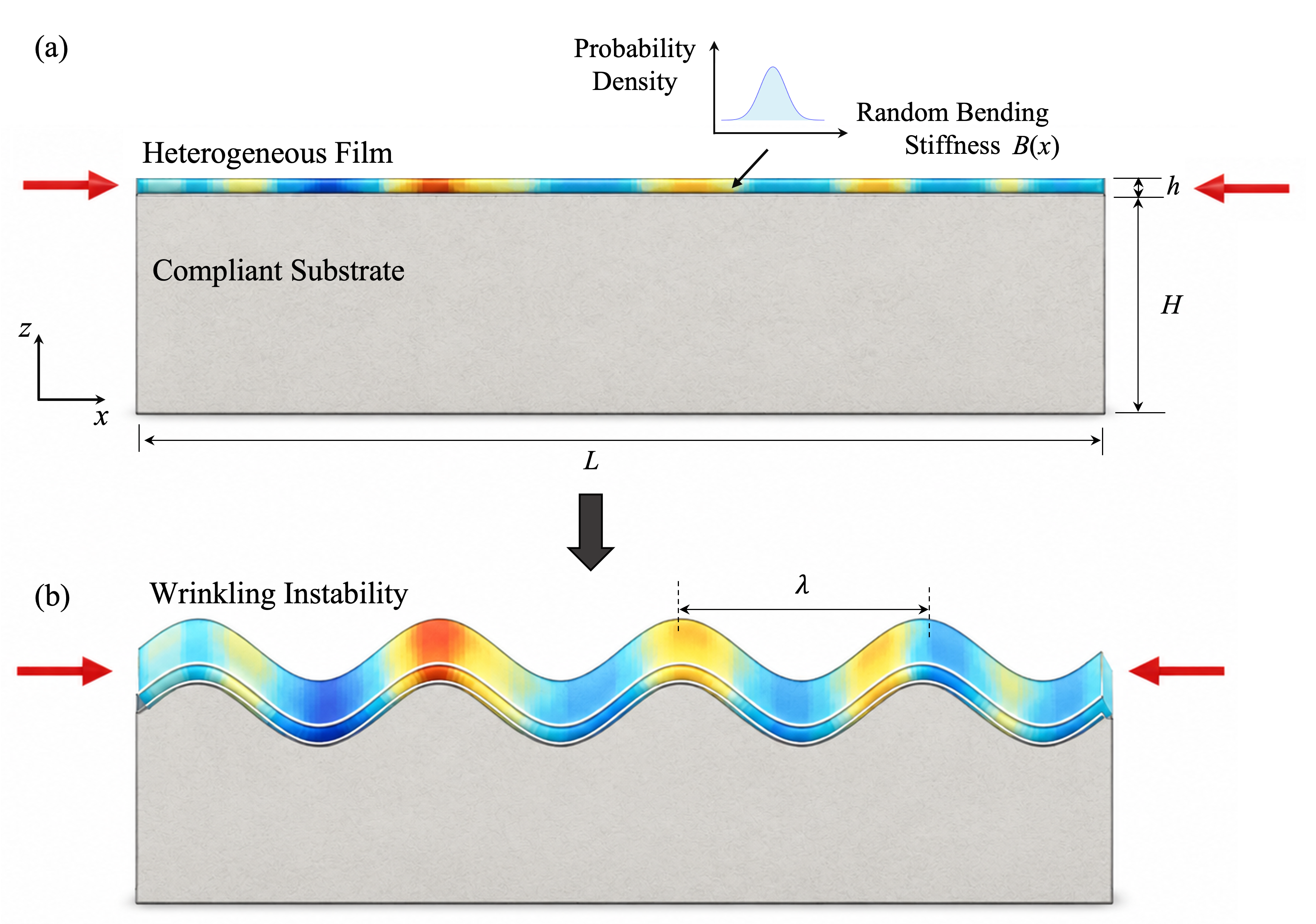}
    \caption{Schematic of the one-dimensional film--substrate system considered in this work.  A stiff film of length $L$ and thickness $h$ is bonded to a compliant substrate and subjected to an in-plane compressive membrane force along the $x$ direction. The out-of-plane deflection of the film is denoted by $w(x)$ in the $z$ direction. The film bending stiffness $B(x)=\bar{E}(x)h^3/12$ is spatially heterogeneous and treated as a random field, whereas the substrate is assumed to be homogeneous and represented by a linear normal reaction operator $p=\mathcal{K}[w]$, or equivalently by the Fourier-domain substrate kernel $K(q)$. }
    \label{fig:film_substrate}
\end{figure}

Consider a hard film of length $L$ and thickness $h$ ($h\ll L$) bonded to a soft substrate surface, as illustrated in Fig.~\ref{fig:film_substrate}.
The film is subjected to in-plane compression along the axial direction, i.e., $x-$ direction, resulting in a deflection $w(x)$ in the $z$ direction.
The film is assumed to be linearly elastic, with spatially varying Young's modulus $E(x)$ and constant Poisson's ratio $\nu$. Under the plane-strain approximation, the effective modulus and bending stiffness are
\begin{equation}
\bar{E}(x)=\frac{E(x)}{1-\nu^2},
\qquad
B(x)=\frac{\bar{E}(x)h^3}{12}.
\label{eq:bending_stiffness_def}
\end{equation}
The spatial variation of $B(x)$ represents the random bending-stiffness heterogeneity considered in this work.

The substrate exerts a normal reaction force $p(x)$ on the thin film, which is expressed as
\begin{equation}
p(x)=\mathcal{K}[w](x),
\label{eq:substrate_operator}
\end{equation}
where $\mathcal{K}$ is the linear response operator of the substrate.
We assume that the substrate is homogeneous in this work.
Under periodic or infinite-domain conditions, $\mathcal{K}$ can be diagonalized in Fourier space, yielding a scalar relation between the Fourier transforms of the normal reaction force and the deflection:
\begin{equation}
\widehat{p}(q)=K(q)\,\widehat{w}(q),
\label{eq:substrate_kernel}
\end{equation}
where $K(q)$ denotes the substrate stiffness kernel, which generally depends on the wavenumber $q$. 
This unified formulation includes both the constant stiffness $K(q)=K$ corresponding to the liquid/Winkler foundation approximation and the wavenumber-dependent stiffness $K(q)$ for elastic half-spaces.

\subsection{Governing Equations}
\label{subsec:gov-equation}
A one-dimensional von K\'arm\'an nonlinear model is used to describe the response of the film mid-plane.
$u(x)$ denotes the in-plane axial displacement of the film mid-plane, while $w(x)$ is the out-of-plane deflection. The total axial strain in the film is then given by
\begin{equation}
\varepsilon(x)
=\varepsilon^0(x)+\frac{\mathrm{d}u}{\mathrm{d}x}
+\frac{1}{2}\left(\frac{\mathrm{d}w}{\mathrm{d}x}\right)^2.
\label{eq:1d_nonlinear_strain}
\end{equation}
For convenience we use $'$ to denote derivative with respect to $x$.

The in-plane membrane force in the film $N(x)$ and strain satisfy the linear elastic constitutive law:
\begin{equation}
N(x) = h \bar{E}(x)\left[ \varepsilon^0 + u' + \frac{1}{2} (w')^2 \right].
\label{eq:constitutive}
\end{equation}

The substrate exerts tangential and normal tractions $\boldsymbol{T} =(T_1,T_2)$ at the interface, where $T_1$ is the shear stress and $T_2$ is the normal stress, which are governed by the following equilibrium equations:
\begin{align}
\label{eq:inplane_equil_general}
&N'(x)=T_1(x), \\
\label{eq:outplane_equil_general}
&\left(B(x)w''\right)''+\!\left(N\,w'\right)'+T_2(x)=0.
\end{align}

When the substrate is a liquid substrate or a Winkler-type foundation, the interface provides no tangential stress, and the normal stress is locally proportional to the deflection, i.e., $T_1\equiv 0$ and $T_2=K w$,
Consequently, equation \eqref{eq:outplane_equil_general} reduces to a scalar governing equation in terms of $w(x)$:
\begin{equation}
\left(B(x)w''\right)''+N\,w''+Kw=0.
\label{eq:outplane_equil_general_liquid}
\end{equation}

For cases in which shear tractions cannot be neglected, such as a solid substrate, we adopt the formulation for the surface response of an elastic half-space given in \cite{huang2005nonlinear}.
For an isotropic, linearly elastic solid half-space (with Young's modulus $E_s$ and Poisson's ratio $\nu_s$, under the plane-strain and one-dimensional wrinkling assumption ($k_2=0$),
only the tangential displacement $u(x)$ and normal displacement $w(x)$ are retained.
Let the Fourier domain wavenumber be $q\equiv k_1$, and denote $k=|q|$. The interfacial stresses and displacements then satisfy the following relation:
\begin{equation}
\begin{pmatrix}
\hat T_1\\[2pt]
\hat T_2
\end{pmatrix}
=
\begin{pmatrix}
D_{11}(q) & D_{12}(q)\\
D_{21}(q) & D_{22}(q)
\end{pmatrix}
\begin{pmatrix}
\hat u\\[2pt]
\hat w
\end{pmatrix}.
\label{eq:D_matrix_def}
\end{equation}

The explicit one-dimensional forms of $D_{ij}(q)$ are
\begin{align}
D_{11}(q) &= 4C_{\mathrm{sub}}(1-\nu_s)\,|q|,
\label{eq:D11_solid}\\
D_{22}(q) &= 4C_{\mathrm{sub}}(1-\nu_s)\,|q|,
\label{eq:D22_solid}\\
D_{12}(q) &= 2 i\,C_{\mathrm{sub}}(1-2\nu_s)\,q,
\label{eq:D12_solid}\\
D_{21}(q) &= -2 i\,C_{\mathrm{sub}}(1-2\nu_s)\,q,
\label{eq:D21_solid}
\end{align}
with
\begin{equation}
C_{\mathrm{sub}}
=\frac{\bar E_s(1-\nu_s)}{6-8\nu_s}
=\frac{\bar E_s(1-\nu_s)}{2(3-4\nu_s)},
\qquad
\bar E_s=\frac{E_s}{1-\nu_s^2}.
\label{eq:Csub_def}
\end{equation}
Equation \eqref{eq:D_matrix_def}  reflects the tangential–normal coupling at the interface of a solid substrate: $D_{12}$ and $D_{21}$ are non-zero, implying that both $T_1$ and $T_2$ generally depend on $u$ and $w$. Therefore, for the case of a solid half-space, equations 
\eqref{eq:inplane_equil_general}--\eqref{eq:outplane_equil_general} and \eqref{eq:D_matrix_def} must be solved simultaneously as a coupled system for $(u,w)$.

The strong-contrast expansion developed below is formulated for a scalar normal substrate kernel.
For Winkler or liquid substrates, this scalar representation is exact, with $K(q)\equiv K$. 
For an elastic half-space, the coupled interfacial response  is condensed at the linear instability level by eliminating the in-plane displacement.
This condensation allows Winkler/liquid substrates and elastic half-space substrates to be treated within the same scalar dispersion formulation. 
The detailed derivation of $K_{\mathrm{eff}}(q)$ is provided in ~\ref{app:Keff}.
The unified substrate kernel is written as
\begin{equation}
K(q)=
\begin{cases}
K, & \text{Winkler/liquid substrate},\\
K_{\mathrm{eff}}(q), & \text{Elastic half-space substrate}.
\end{cases}
\label{eq:Ks_cases_ch2}
\end{equation}

\subsection{Homogeneous Benchmark}
To start with, we briefly review the main results concerning homogeneous film--substrate system, i.e. the film has a constant bending stiffness ($B(x)\equiv B$). In this case, Eq.~\eqref{eq:outplane_equil_general_liquid} reduces to the classical beam-on-Winkler-foundation equation:
\begin{equation}
  B w'''' + N w'' + K w = 0.
\label{eq:beam_winkler_uniform}
\end{equation}

For an infinitely long system, or a finite system with periodic boundary conditions, we seek a Fourier-mode solution $w(x)=A e^{iqx}$ with $A\neq 0$. Substitution into Eq.~\eqref{eq:beam_winkler_uniform} gives the dispersion relation
$D_0(q;N)=B q^4 - N q^2 + K = 0$,
whose minimum is attained at $q_c^2={N/(2B)}$. Therefore, instability occurs when the minimum value of $D_0(q;N)$ becomes non-positive, yielding the classical results $N_c = 2\sqrt{BK}$,
and $q_c = \left(\frac{K}{B}\right)^{1/4}$.
These expressions provide the benchmark instability criterion for the homogeneous beam-on-Winkler-foundation system and will be used for comparison with the heterogeneous case.

\subsection{Numerical Methods for Heterogeneous Systems}
For heterogeneous systems, the instability threshold can no longer be obtained from a closed-form dispersion relation and must instead be determined numerically. 
In this subsection, we employ two complementary approaches: a generalized eigenvalue method to compute the critical load and mode, and a Fourier spectral iteration scheme to solve the coupled equilibrium equations for solid-substrate systems.

\subsubsection{Generalized Eigenvalue Method}
When the bending stiffness $B(x)$ varies spatially, Eq.~\eqref{eq:outplane_equil_general_liquid} becomes a fourth-order ordinary differential equation with variable coefficients, and the problem no longer admits a scalar dispersion relation in terms of a single wavenumber. The instability condition can instead be recast as the generalized eigenvalue problem
\begin{equation}
\mathcal{M}[w] = N\,\mathcal{Q}[w],
\label{eq:generalized_eigen}
\end{equation}
with

\begin{equation}
\mathcal{M}[w]
=
\left(B(x) w_{,xx}\right)_{,xx}
+
\mathcal K[w],
\qquad
\mathcal{Q}[w] = - w_{,xx},
\label{eq:operators}
\end{equation}
and
\begin{equation}
\widehat{\mathcal K[w]}(q)=K(q)\hat w(q).
\label{eq:K_operator_fourier}
\end{equation}
The smallest positive eigenvalue gives the critical load, and the corresponding eigenvector gives the critical mode.

Numerically, Eq.~\eqref{eq:generalized_eigen} is discretized using a Fourier pseudospectral method, in which spatial derivatives are evaluated in Fourier space and variable-coefficient terms are treated in physical space through FFT-based transforms. 
A standard $3/2$-rule de-aliasing procedure is used to reduce aliasing errors arising from the variable-coefficient term. 
The smallest positive eigenvalue is computed using the locally optimal block preconditioned conjugate gradient (LOBPCG) method, and the associated critical wavenumber is identified from the dominant peak of the Fourier amplitude spectrum $|\hat{w}(q)|$.

\subsubsection{Fourier Spectral Method}
To compute the critical configuration of a heterogeneous film--substrate system under a prescribed external load, we adopt a Fourier-space fixed-point iteration scheme inspired by the work of~\cite{huang2005nonlinear}. 
Based on the in-plane and out-of-plane equilibrium equations, Eqs.~\eqref{eq:inplane_equil_general}--\eqref{eq:outplane_equil_general}, together with the interfacial displacement--traction relation for a solid half-space, Eq.~\eqref{eq:D_matrix_def}, the coupled governing equations are projected into Fourier space and solved iteratively.

We define the Fourier transform of the out-of-plane displacement as  
\begin{equation}
\hat{w}(q)=\mathcal{F}[w](q)=\int_{-\infty}^{\infty} w(x)e^{-iqx}\,\mathrm{d}x ,
\label{eq:fourier_def}
\end{equation}
and similarly denote the Fourier transform of the in-plane displacement $u(x)$ by $\hat{u}(q)$. To describe random heterogeneity in the film stiffness, we decompose
\begin{equation}
B(x)=B_0+\Delta B(x),\qquad B_0=\langle B\rangle,
\label{eq:B0deltaB}
\end{equation}
where $B_0$ is the reference bending stiffness. The corresponding reference in-plane stiffness is defined by
\begin{equation}
h\bar E(x)=\frac{12}{h^2}B(x),
\qquad
h\bar E_0=\frac{12}{h^2}B_0,
\end{equation}
with $\bar E_0=E_0/(1-\nu^2)$ and $B_0=\bar E_0 h^3/12$.
Applying the Fourier transform to Eq.~\eqref{eq:outplane_equil_general} and using Eq.~\eqref{eq:D_matrix_def}, we obtain
\begin{equation}
\Bigl(D_{22}(q)+B_0 q^4-N_0 q^2\Bigr)\,\hat w(q)
=
-\,D_{21}(q)\,\hat u(q)
-\,q^2\,\mathcal F\!\Big[\Delta B(x)\,w_{xx}(x)\Big](q)
-\,i q\,\mathcal F\!\Big[ N(x)\,w_x(x)\Big](q).
\label{eq:w_update_solid_cn}
\end{equation}
Here, the left-hand side of Eq.~\eqref{eq:w_update_solid_cn} contains only the contribution from the homogeneous reference medium, while the heterogeneity-induced convolution terms are treated explicitly on the right-hand side. 
To improve iterative stability, a numerical viscosity $\zeta_w$ is introduced, leading to the update
\begin{equation}
\hat w^{\,n+1}
=
\Bigl(D_{22}(q)+B_0 q^4+\zeta_w\Bigr)^{-1}
\Bigl\{
q^2\,\mathcal F\!\big[\Delta B(x)\,w_{xx}^{\,n}(x)\big]
- i q\,\mathcal F\!\big[ N(x)^{n}\,w_x^{n}(x)\big]
- D_{21}(q)\,\hat u^{\,n}
+\zeta_w\,\hat w^{\,n}
\Bigr\},
\label{eq:w_update_viscous}
\end{equation}

Similarly, the in-plane displacement is updated from the in-plane equilibrium equation as
\begin{equation}
\hat u^{\,n+1}
=
\Bigl(D_{11}(q)+h\bar E_0\,q^2+\zeta_u\Bigr)^{-1}
\Bigl\{
- D_{12}(q)\,\hat w^{\,n+1}
+ i q\,\mathcal F\!\big[\Psi_u^{\,n}(x)\big]
+\zeta_u\,\hat u^{\,n}
\Bigr\},
\label{eq:u_update_viscous}
\end{equation}
where $\zeta_u$ is another damping parameter and
\begin{equation}
\Psi_u^{\,n}(x)
:=h\,\Delta\bar E(x)\,u_x^{\,n}(x)
+h\,\bar E(x)\Bigl[\varepsilon^0(x)+\tfrac12\bigl(w_x^{\,n+1}(x)\bigr)^2\Bigr].
\label{eq:Psiu_def_viscous}
\end{equation}

\section{Strong Contrast Expansion (SCE) and Homogenized Instability Criterion}
\label{sec:1dsce}

For a one-dimensional thin beam model with a random stiffness field $B(x)$, 
the statistical properties of the stiffness fluctuations cause the critical instability load $N_c$ and the critical wavenumber $q_c$ to deviate from their homogeneous counterparts. 
In the homogeneous case, the instability behavior of a thin beam on an elastic foundation is described by Eq.~\eqref{eq:beam_winkler_uniform}.
When spatial heterogeneity is present in the material, the governing equation becomes $(B(x) w'')'' + N w'' + K w = 0$.
In this case, different wavenumbers interact through the convolution coupling introduced by $B(x)$ and the system can no longer be characterized by a simple algebraic dispersion relation for a single wavenumber.

To obtain a homogenized criterion for predicting the instability threshold, we seek an effective dispersion relation $D^{\rm eff}(q,N)=0$ that incorporates the influence of random stiffness fluctuations,
from which the effective critical load $N_c^{\rm eff}$ and effective critical wavenumber $q_c^{\rm eff}$ for the random medium can be derived.
In addition, we further introduce below a scalar self-consistent kernel update for the load parameter entering the cavity kernel, which improves the numerical prediction of the critical pair while preserving the same second- and third-order SCE truncation.

\subsection{Lippmann--Schwinger Equation and Cavity Field Decomposition}
Consider the linear governing equation for a beam with spatially varying bending stiffness on a linear, translationally invariant substrate. In Fourier space, the substrate response is represented by the scalar kernel \(K(q)\). The Fourier-space Green's function associated with the reference homogeneous medium is defined by
\begin{equation}
D_0(q;N)=B_0 q^4-Nq^2+K(q),
\label{eq:D0_unified}
\end{equation}
and
\begin{equation}
\widehat{G}_0(q)=\frac{1}{D_0(q;N)}
=
\frac{1}{B_0 q^4-Nq^2+K(q)}.
\label{eq:G0hat}
\end{equation}
For a Winkler or liquid substrate, \(K(q)\equiv K\). For an elastic half-space substrate, \(K(q)=K_{\mathrm{eff}}(q)\) after the condensation procedure described in ~\ref{app:Keff}.

By introducing the curvature $\kappa(x)=w''(x)$ and the polarization $\tau(x)=\Delta B(x)\,\kappa(x)$, Eq.~\eqref{eq:outplane_equil_general_liquid} can be equivalently reformulated into a convolution form for $\kappa$
\begin{equation}
\kappa = -\,\Xi^0 \ast \tau,
\qquad
\widehat{\Xi^0}(q)=\frac{q^4}{D_0(q;N)}.
\label{eq:kappa_LS_main}
\end{equation}
where $*$ denotes the convolution operator. The detailed derivation of Eq.~\eqref{eq:kappa_LS_main} is provided in~\ref{app:LS-kappa}. 
Note that $\widehat{\Xi^0}(q)$ tends to a constant as $|q| \to \infty$, therefore, the kernel can be decomposed into a local part and a nonlocal part
\begin{equation}
\Xi^0(x)=D^0\,\delta(x)+H^0(x),
\qquad
D^0=\frac{1}{B_0},
\label{eq:Xi_split_main}
\end{equation}
and the cavity field for curvature is defined by
\begin{equation}
\mathcal F_\kappa = -\,H^0\ast \tau,
\qquad
\kappa = \mathcal F_\kappa - D^0\,\tau.
\label{eq:cavity_def_main}
\end{equation}

This leads to a local relation
\begin{equation}
\tau = L_B(x)\,\mathcal F_\kappa,
\qquad
L_B(x)=\frac{\Delta B(x)}{1+D^0\Delta B(x)}.
\label{eq:LB_main}
\end{equation}
Further details on the kernel decomposition and cavity-field reformulation  are presented in~\ref{app:cavity}.

Defining $\mathcal H:=-H^0$, the cavity field satisfies
\begin{equation}
\mathcal F_\kappa = \mathcal F_{\rm ext} + \mathcal H\ast \tau,
\label{eq:cavity_eq_main}
\end{equation}
where $\mathcal F_{\rm ext}$ is an external field introduced for formal closure, and the limit $\mathcal F_{\rm ext}\to 0$ is taken in the eigenvalue problem.

\subsection{Strong Contrast Expansion: Effective Susceptibility $L_e$}

From Eqs.~\eqref{eq:LB_main} and~\eqref{eq:cavity_eq_main} the microscopic response operator can be formally defined as
\begin{equation}
S = L\bigl[I-\mathcal H L\bigr]^{-1}
= L + L\mathcal HL + L\mathcal H L\mathcal HL + \cdots,
\label{eq:S_def_main}
\end{equation}
where $L$ denotes the multiplicative operator $L_B(x)$. 
After the ensemble averaging of~\eqref{eq:S_def_main} and reorganizing the connected clusters, the effective susceptibility operator $L_e$ is defined through
\begin{equation}
\langle \tau\rangle = L_e\,\langle \mathcal F_\kappa\rangle.
\label{eq:Le_def_main}
\end{equation}

We decompose $L=a+\delta L$, where $a=\langle L\rangle$ and $\langle\delta L\rangle=0$.
For the practical third-order prediction used in this work, we evaluate the cavity kernel entering the connected-cluster terms with an auxiliary kernel load parameter $N_{\rm ker}$
\begin{equation}
\widehat{\mathcal H}(q;N_{\rm ker})
=
\frac{K(q)-N_{\rm ker}q^2}
{B_0\left[B_0 q^4-N_{\rm ker}q^2+K(q)\right]}.
\label{eq:H_kernel_sc_main}
\end{equation}
Accordingly, the effective susceptibility is written as $L_e(q;N_{\rm ker})$. 
Here, $N_{\rm ker}$ is not prescribed a priori, but will be determined below through a scalar self-consistent iteration together with the critical load and critical wavenumber.
Truncated to $\mathcal O(\mathcal H^2)$, i.e., connected clusters up to third order, we have
\begin{equation}
L_e(q;N_{\rm ker}) = a + \Delta_2(q;N_{\rm ker}) + \Delta_3(q;N_{\rm ker}) + \mathcal O(\mathcal H^3).
\label{eq:Le_3rd_main}
\end{equation}
where the second- and third-order connected terms in spectral space are expressed as
\begin{align}
\Delta_2(q;N_{\rm ker}) &= \widehat{\mathcal H}(q;N_{\rm ker})\,\widehat{W_2}(q),
\label{eq:Delta2_spec_main}\\
\Delta_3(q;N_{\rm ker}) &= \int_{\mathbb R}\frac{{\rm d}p}{2\pi}\;
\widehat{\mathcal H}(p;N_{\rm ker})\,\widehat{\mathcal H}(q-p;N_{\rm ker})\,\widehat{W_3}(p,q-p),
\label{eq:Delta3_spec_main}
\end{align}
with
\[
W_2(r)=\langle \delta L(0)\delta L(r)\rangle,
\qquad
W_3(r_1,r_2)=\langle \delta L(0)\delta L(r_1)\delta L(r_2)\rangle_{\rm conn}.
\]

The derivation of the connected clusters including the expansion of $\langle S\rangle^{-1}$ is detailed in~\ref{app:Le3-proof}.

\subsection{Dyson-Type Effective Dispersion Relation, Critical Criterion, and Self-Consistent Kernel Update}

In the eigen-problem limit with no external field ($\mathcal F_{\rm ext}\to 0$), the closure condition becomes
\begin{equation}
1-\widehat{\mathcal H}(q;N)\,L_e(q;N_{\rm ker})=0.
\label{eq:closure_main}
\end{equation}

This can be written as a Dyson-type effective dispersion relation
\begin{equation}
D_{\rm eff}(q;N\,|\,N_{\rm ker})
=
D_0(q;N)+\Sigma_{\rm cav}(q;N\,|\,N_{\rm ker})=0,
\label{eq:Dyson_main_a}
\end{equation}
with
\begin{equation}
\Sigma_{\rm cav}(q;N\,|\,N_{\rm ker})
=
\frac{Nq^2-K(q)}{B_0}\,L_e(q;N_{\rm ker}).
\label{eq:Dyson_main}
\end{equation}

For a prescribed kernel load $N_{\rm ker}$, the SCE truncation is first used to evaluate $L_e(q;N_{\rm ker})$, while the physical load $N$ entering the instability criterion is then determined from the closure equation itself. Defining $\phi(q;N_{\rm ker}) := 1-\frac{L_e(q;N_{\rm ker})}{B_0}$, Eq.~\eqref{eq:closure_main} is equivalently rewritten as
\begin{equation}
N(q;N_{\rm ker})
=
\frac{B_0 q^4 + K(q)\,\phi(q;N_{\rm ker})}
     {q^2\,\phi(q;N_{\rm ker})},
\qquad q>0,\ \phi(q;N_{\rm ker})>0.
\label{eq:N_of_q_main}
\end{equation}

For a fixed trial value of $N_{\rm ker}$, the critical pair is obtained by minimizing $N(q;N_{\rm ker})$ over admissible positive wavenumbers. 
Equivalently, the stationary condition can be written as
\begin{equation}
2B_0 q^4\,\phi(q;N_{\rm ker})
+
q^5\,\partial_q L_e(q;N_{\rm ker})
+
\phi(q;N_{\rm ker})^2
\left[
qK'(q)-2K(q)
\right]
=0 .
\label{eq:stationary_sc_main}
\end{equation}

To improve the prediction accuracy while retaining the same second- and third-order SCE truncation, we employ a scalar fixed-point self-consistent update for the kernel load. 
Starting from  an initial guess $N_{\rm ker}^{(0)} = 0$, we compute at the $n$-th step the trial critical pair $\bigl(q_c^{(n)},N_c^{(n)}\bigr)$ from Eq.~\eqref{eq:N_of_q_main} (or equivalently from Eq.~\eqref{eq:stationary_sc_main}), and then update the kernel load according to
\begin{equation}
N_{\rm ker}^{(n+1)}
=
(1-\theta)\,N_{\rm ker}^{(n)}
+
\theta\,N_c^{(n)},
\qquad 0<\theta\le 1.
\label{eq:self_consistent_update_main}
\end{equation}

In the present numerical implementation we take $\theta=1$, so that the updated kernel load is directly set to the current critical-load prediction. 
The iteration is terminated when both $N_{\rm ker}$ and $q_c$ become numerically converged, and the converged values are denoted by
\begin{equation}
N_c^{\rm sc} := \lim_{n\to\infty} N_c^{(n)},
\qquad
q_c^{\rm sc} := \lim_{n\to\infty} q_c^{(n)}.
\label{eq:sc_limit_main}
\end{equation}

Numerically, the third-order SCE prediction is obtained by computing the spectral forms of $\Delta_2$ and $\Delta_3$ in Equations~\eqref{eq:Delta2_spec_main}--\eqref{eq:Delta3_spec_main}, evaluating $L_e(q;N_{\rm ker})$, and then performing the above self-consistent kernel-load iteration until convergence.

\section{Gaussian Random Field Generation and Computation of $W_2,W_3$}
\label{sec:random_W2W3}
This section defines the random bending-stiffness field used in the homogenized instability analysis and derives the two- and three-point connected statistics entering the second- and third-order SCE corrections. Numerical details of the random-field generation are provided in ~\ref{app:numerics}.

\subsection{Standard Gaussian Random Field $g(x)$ }
\label{subsec:GRF_model}
We consider a zero-mean, unit-variance stationary Gaussian random field $g(x)$ defined on a periodic interval $[0,L)$. Its spatial correlation structure is prescribed through a target power spectral density (PSD). 
To represent both low-wavenumber modulation and high-wavenumber cutoff, we adopt the power-law-modulated Gaussian spectrum

\begin{equation}
S_{\alpha}^{\mathrm{raw}}(k)
=\left(|k|+k_0\right)^{\alpha}
\exp\!\left[-\frac{k^2}{2\sigma_k^2}\right],
\label{eq:psd_raw_main}
\end{equation}
which is then normalized in the discrete sense to obtain the target spectrum $S_g(k_r)$ used for numerical sampling.
Here, $k_0>0$ regularizes the spectrum near $k=0$, $\sigma_k$ controls the spectral bandwidth, and $\alpha$ adjusts the distribution of spectral weight across wavenumbers.

The dominant material wavenumber is defined as the peak position of 
Eq.~\eqref{eq:psd_raw_main}. For $\alpha>0$, it is given by
\begin{equation}
k^\ast=
\frac{-k_0+\sqrt{k_0^2+4\alpha\sigma_k^2}}{2}.
\label{eq:kstar_def}
\end{equation}
The corresponding dominant material wavelength is defined as
\begin{equation}
\lambda^\ast=\frac{2\pi}{k^\ast}.
\label{eq:lambda_star_def}
\end{equation}
This length scale will be used below to characterize the interaction between the material heterogeneity and the selected wrinkling wavelength.

\subsection{Exponential Mapping to the Stiffness Field $B(x)$}
To ensure strict positivity of the bending stiffness and allow adjustable contrast, we introduce the exponential mapping
\begin{equation}
B(x)=B_m\exp\!\Big(\varepsilon\,g(x)-\tfrac{1}{2}\varepsilon^2\Big),
\label{eq:lognormal_B_main}
\end{equation}
which generates a log-normal random field for the stiffness. 
Here, $B_m$ is the nominal mean stiffness and $\varepsilon$ is a dimensionless contrast parameter controlling the fluctuation amplitude. 
The term $-\tfrac{1}{2}\varepsilon^2$ ensures that the theoretical mean satisfies $\langle B(x)\rangle=B_m$.

\begin{figure}[H]
    \centering
    \includegraphics[width=1.0\textwidth]{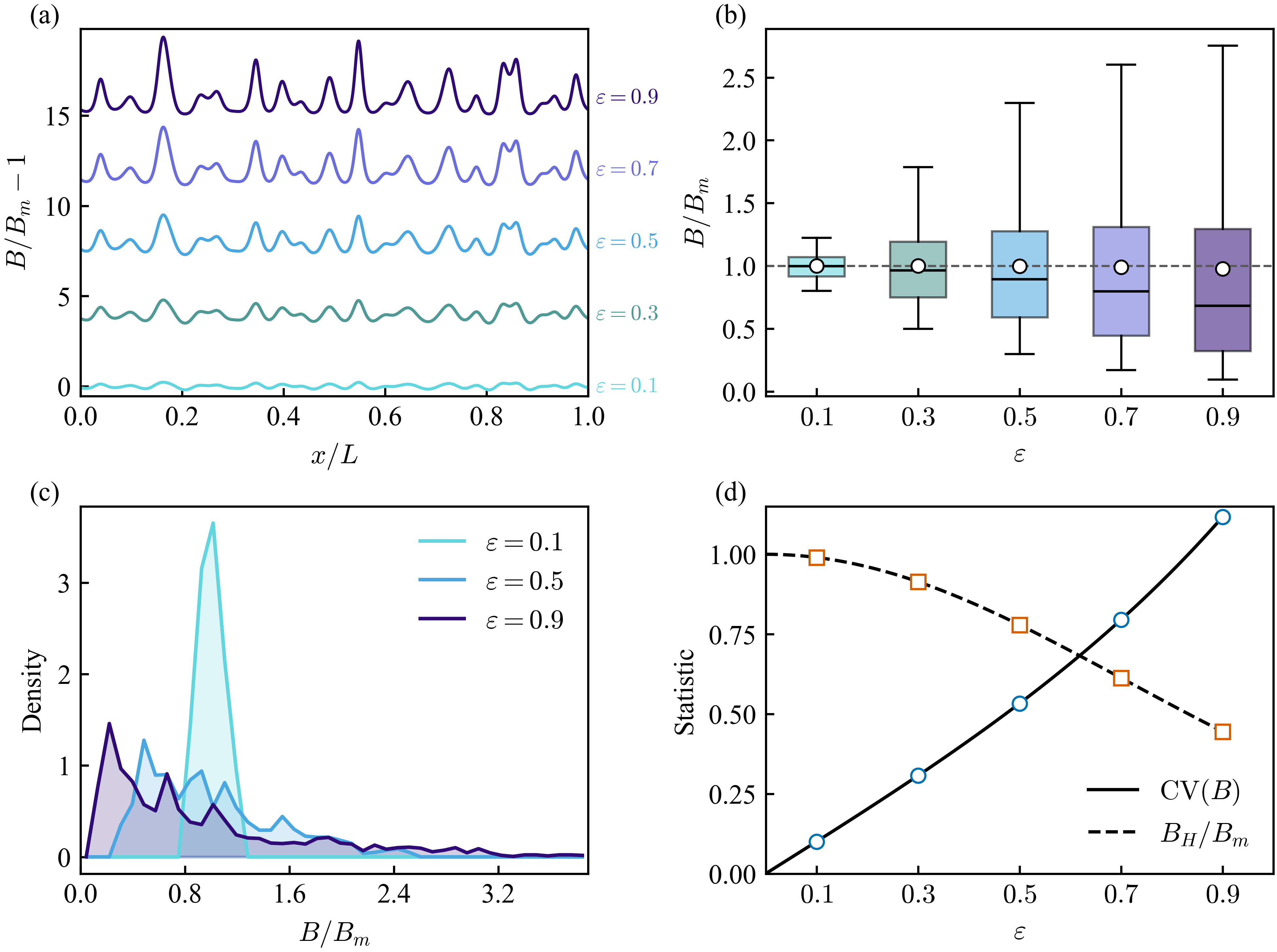}
    \caption{ Effect of the contrast parameter $\varepsilon$ on the randomly heterogeneous bending-stiffness field $B(x)$ generated by the log-normal mapping  $B(x)=B_m\exp[\varepsilon g(x)-\varepsilon^2/2]$. 
    The same underlying zero-mean, unit-variance Gaussian random field $g(x)$ is used for all values of $\varepsilon$, so that changes in the plotted fields arise from the stiffness contrast rather than from changes in the spatial correlation pattern.
    The spectral parameters are fixed as $\alpha=5$, $\sigma_k=2.0$, and $k_0=0.05$, with $B_m=1$.
    (a) Representative normalized stiffness fields $B/B_m-1$ for $\varepsilon=0.1,0.3,0.5,0.7,$ and $0.9$, vertically shifted for clarity.
    Increasing $\varepsilon$ amplifies the stiffness fluctuations while preserving the main peak and valley locations.
    (b) Box plots of the normalized stiffness $B/B_m$ sampled over the spatial domain.
    Boxes indicate the interquartile range, black horizontal lines denote medians, whiskers show the data range within $1.5$ times the interquartile range, and white circles denote sample means.
    The gray dashed line marks the nominal mean level $B/B_m=1$.
    (c) Empirical probability density distributions of $B/B_m$ for representative contrasts $\varepsilon=0.1,0.5,$ and $0.9$.
    Larger $\varepsilon$ broadens the distribution and produces a more pronounced right-skewed tail, reflecting the log-normal nature of the stiffness field.
    (d) Theoretical variation of the coefficient of variation $\mathrm{CV}(B)=\sqrt{\exp(\varepsilon^2)-1}$ and the normalized harmonic-mean stiffness $B_H/B_m=\exp(-\varepsilon^2)$ with $\varepsilon$.
    The markers correspond to the contrast values used in panels (a) and (b).
    }
    \label{fig:GRF_contrast_effect}
\end{figure}

Fig.~\ref{fig:GRF_contrast_effect} illustrates how the contrast parameter $\varepsilon$ controls the statistical heterogeneity of the random bending-stiffness field generated by the log-normal mapping in Eq.~\eqref{eq:lognormal_B_main}. 
In this figure, all stiffness fields are generated from the same underlying Gaussian random field $g(x)$, while only $\varepsilon$ is varied. 
Therefore, the comparison isolates the effect of stiffness contrast from that of the spatial correlation structure. 
As shown in Fig.~\ref{fig:GRF_contrast_effect}(a), increasing $\varepsilon$ amplifies the fluctuation amplitude of $B/B_m-1$, whereas the main peak and valley locations remain almost unchanged. 
This confirms that $\varepsilon$ mainly changes the intensity of the stiffness fluctuations rather than the dominant spatial pattern of the random field.

The statistical distribution of the stiffness values is further quantified in Fig.~\ref{fig:GRF_contrast_effect}(b) and (c). 
For each value of $\varepsilon$, the box plot in Fig.~\ref{fig:GRF_contrast_effect}(b) is constructed from the discrete spatial samples $\{B(x_i)/B_m\}_{i=1}^{N}$ of the corresponding stiffness realization. 
Thus, each box summarizes the distribution of normalized stiffness values over the entire spatial domain. 
With increasing $\varepsilon$, both the interquartile range and the whisker length increase, indicating a broader stiffness distribution and stronger material heterogeneity. 
The white circles remain close to the reference level $B/B_m=1$, consistent with the normalization of the log-normal mapping, while the medians shift downward as the distribution becomes increasingly right-skewed.
Fig.~\ref{fig:GRF_contrast_effect}(c) shows the empirical probability density functions of $B/B_m$ for representative values of $\varepsilon$. 
These density curves are estimated from the same spatial samples $\{B(x_i)/B_m\}_{i=1}^{N}$ using normalized histograms. 
For small contrast, the distribution is sharply concentrated around $B/B_m\approx 1$. 
As $\varepsilon$ increases, the distribution becomes wider and develops a long right tail, which is characteristic of a log-normal random field. 
This implies that strong contrast produces many relatively soft regions together with a small number of very stiff regions.

The consequences of this distributional broadening are summarized by the two theoretical statistics in Fig.~\ref{fig:GRF_contrast_effect}(d). 
The coefficient of variation,
$\mathrm{CV}(B)=\sqrt{\exp(\varepsilon^2)-1}$, increases monotonically with $\varepsilon$, confirming the growth of relative stiffness fluctuations. 
In contrast, the normalized harmonic-mean stiffness,
$B_H/B_m=\exp(-\varepsilon^2)$, decreases monotonically. 
Since the harmonic mean is more sensitive to low-stiffness regions, this decrease indicates that increasing random contrast effectively softens the film from the viewpoint of bending-dominated instability. 
This statistical softening provides a physical basis for the reduction of the critical wrinkling load and the shift of the unstable mode observed in the subsequent analysis.

\begin{figure}[H]
    \centering
    \includegraphics[width=1.0\textwidth]{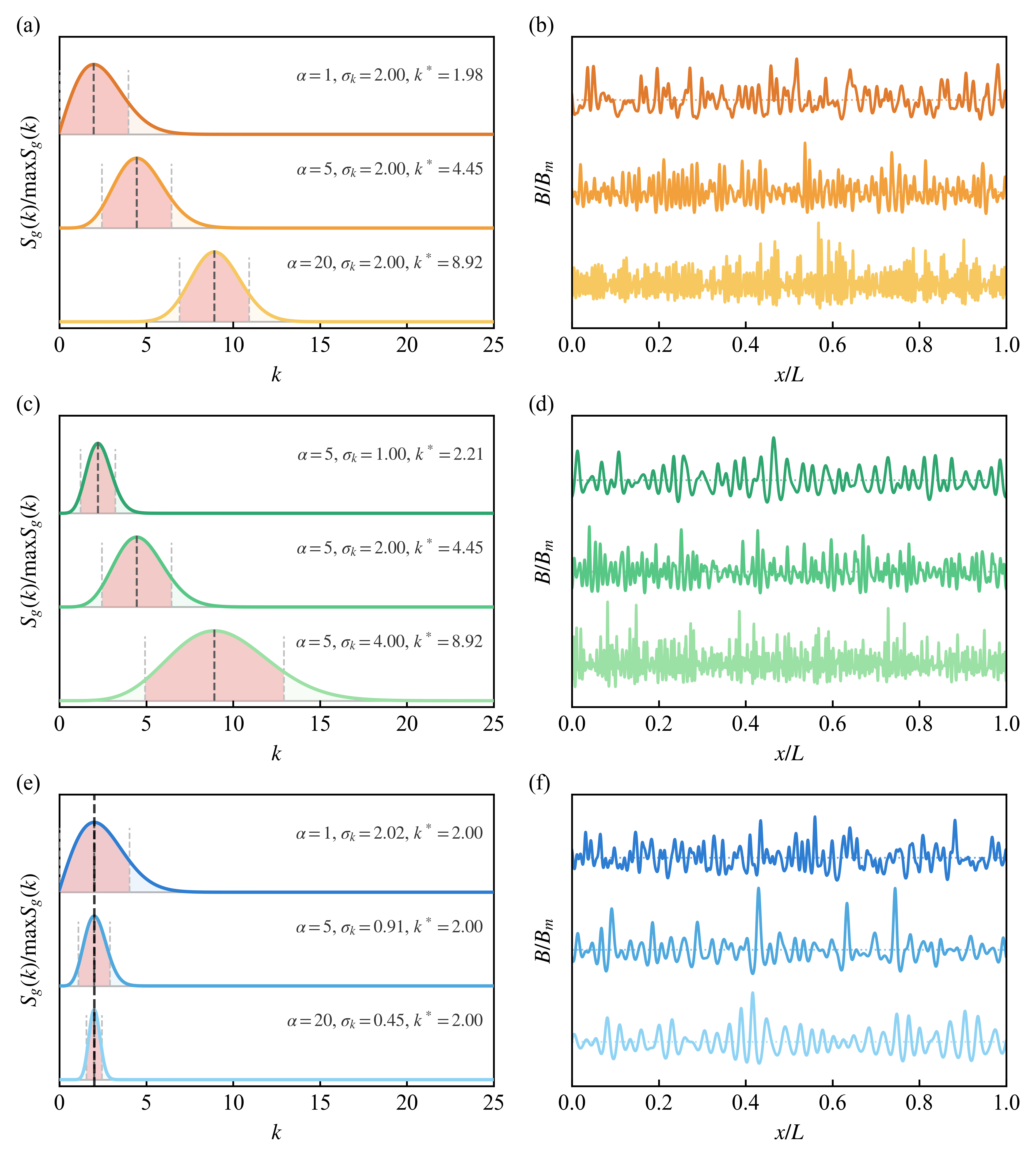}
    \caption{Effects of spectral parameters on the target power spectral density and the corresponding random bending-stiffness fields.
    The left column shows the normalized power spectral density $S_g(k)/\max S_g(k)$ used to generate the zero-mean, unit-variance Gaussian random field $g(x)$.
    The right column shows representative stiffness fields $B(x)$ obtained from $g(x)$ through the exponential mapping $B(x)=B_m\exp[\varepsilon g(x)-\varepsilon^2/2]$.
    All curves are vertically shifted for clarity.
    Unless otherwise specified, $k_0=0.05$, $B_m=1$, and $\varepsilon=0.3$.
    The dashed vertical line in each spectrum denotes the dominant material wavenumber $k^\ast$, and the shaded region highlights the main spectral band around the peak.
    (a,b) Effect of the spectral exponent $\alpha$ at fixed spectral width $\sigma_k=2.0$.
    Increasing $\alpha$ shifts the spectral peak toward larger wavenumbers and therefore reduces the dominant material wavelength $\lambda^\ast=2\pi/k^\ast$.
    (c,d) Effect of the spectral width $\sigma_k$ at fixed spectral exponent $\alpha=5$.
    Increasing $\sigma_k$ broadens the spectrum, weakens the high-wavenumber cutoff, and produces shorter-scale stiffness fluctuations.
    (e,f) Controlled comparison with fixed dominant wavenumber $k^\ast=2.0$.
    Here $\sigma_k$ is adjusted for each $\alpha$ so that the dominant material wavelength remains unchanged, while the spectral shape and local fluctuation structure vary.
    }
    \label{fig:GRF_spectral_parameters}
\end{figure}

Fig.~\ref{fig:GRF_spectral_parameters} further shows how the spectral parameters determine the spatial structure of the random bending-stiffness field.
The left column presents the normalized target power spectral density $S_g(k)/\max S_g(k)$, while the right column shows representative stiffness realizations $B(x)$ obtained from the corresponding Gaussian random fields through the exponential mapping in Eq.~\eqref{eq:lognormal_B_main}.
The dashed vertical line in each spectrum marks the dominant material wavenumber $k^\ast$, and the associated dominant material wavelength is $\lambda^\ast=2\pi/k^\ast$.
The curves in each panel are vertically shifted only for visual clarity.

Figs.~\ref{fig:GRF_spectral_parameters}(a,b) examine the effect of the spectral exponent $\alpha$ at fixed spectral width $\sigma_k=2.0$.
As $\alpha$ increases from $1$ to $20$, the peak of the spectrum shifts from low to high wavenumbers, with $k^\ast$ increasing accordingly.
Therefore, the dominant material wavelength $\lambda^\ast$ decreases.
Fig.~\ref{fig:GRF_spectral_parameters}(b) shows the same trend: small $\alpha$ produces relatively long-wavelength stiffness variations, whereas large $\alpha$ introduces denser fluctuations with shorter spatial scales.
Thus, when $\sigma_k$ is fixed, changing $\alpha$ modifies both the spectral shape and the dominant material length scale.

Figs.~\ref{fig:GRF_spectral_parameters}(c,d) show the effect of the spectral width $\sigma_k$ at fixed $\alpha=5$.
Increasing $\sigma_k$ weakens the high-wavenumber cutoff and shifts the spectral peak toward larger $k$.
As a result, $k^\ast$ increases and the dominant material wavelength decreases.
The corresponding stiffness fields become progressively more oscillatory as $\sigma_k$ increases, indicating that $\sigma_k$ controls the correlation length and the amount of short-wavelength content in the random field.
Figs.~\ref{fig:GRF_spectral_parameters}(e,f) provide a controlled comparison in which the dominant material wavenumber is fixed at $k^\ast=2.0$.
For different $\alpha=1,5,20$, the spectral width $\sigma_k$ is adjusted so that the dominant material wavelength $\lambda^\ast=2\pi/k^\ast$ remains the same.
Although the spectra still differ in shape and bandwidth, and the corresponding stiffness fields show different local fluctuation details, their dominant spatial scale is controlled at the same level.
This comparison shows that $\lambda^\ast$ can be used as a key parameter to characterize the dominant material length scale, while $\alpha$ mainly affects the spectral shape and local fluctuation structure.

\subsection{Closed-form Results for $W_2,W_3$ under Exponential Mapping}

Let $g(x)$ be a stationary Gaussian field with zero mean, unit variance, and covariance function
\begin{equation}
C(r)=\langle g(0)g(r)\rangle .
\end{equation}
We adopt the exponential mapping in Eq.~\eqref{eq:lognormal_B_main} and take the reference stiffness as $B_0=B_m$.
The corresponding local fractional transformation $L=B_0(1-B_0/B)$ yields
\begin{equation}
L(x)=B_m\Bigl[1-\exp\!\bigl(-\varepsilon g(x)+\tfrac12\varepsilon^2\bigr)\Bigr]
= B_m\bigl(1-\beta^2 Y(x)\bigr),
\label{eq:L_from_g_main}
\end{equation}
where $\beta=e^{\varepsilon^2/2},\;\; Y(x)=\exp\!\Bigl(-\varepsilon g(x)-\tfrac12\varepsilon^2\Bigr)$.
Hence
\begin{equation}
a=\langle L\rangle=B_m(1-\beta^2),
\qquad
\delta L(x)=L(x)-a=B_m\beta^2(1-Y(x)).
\end{equation}

Denoting $r=x_2-x_1$, the two-point statistic is
\begin{equation}
W_2(r):=\langle \delta L(0)\delta L(r)\rangle
= B_m^2\beta^4\Bigl(\langle Y(0)Y(r)\rangle-1\Bigr)
= B_m^2\beta^4\Bigl(e^{\varepsilon^2 C(r)}-1\Bigr).
\label{eq:W2_closed_main}
\end{equation}

Similarly, denoting $r_{12}=r_1-r_2$, the third-order cumulant
$W_3(r_1,r_2):=\langle \delta L(0)\delta L(r_1)\delta L(r_2)\rangle_{\rm conn}$
can be written as
\begin{equation}
W_3(r_1,r_2)
= -\,B_m^3\beta^6\Bigl(
e^{\varepsilon^2\,[C(r_1)+C(r_2)+C(r_{12})]}
- e^{\varepsilon^2 C(r_1)}-e^{\varepsilon^2 C(r_2)}-e^{\varepsilon^2 C(r_{12})}
+2\Bigr),
\label{eq:W3_closed_main}
\end{equation}

Applying the Fourier transform to \eqref{eq:W2_closed_main}--\eqref{eq:W3_closed_main}, we can obtain 
$\widehat{W_2}(q)$ and $\widehat{W_3}(p,q-p)$, which can then be substituted into
Eqs.~\eqref{eq:Delta2_spec_main}--\eqref{eq:Delta3_spec_main} to compute
$\Delta_2,\Delta_3$.
The derivation is given in~\ref{app:W2W3-proof}.

\section{Results and Discussion}
\label{sec:results_discussion}
This section presents the main theoretical and numerical results. 
The discussion proceeds from the homogeneous benchmark to randomly heterogeneous films with increasing stiffness contrast and varying material length scales. 
We first validate the numerical implementation, then compare the SCE predictions with generalized eigenvalue calculations, and finally analyze how heterogeneity modifies wavelength selection, spectral complexity, and spatial localization of the critical mode.

\subsection{Homogeneous Benchmark and Reference Wrinkling Wavelength}
\label{subsec:benchmark_homogeneous}

We first consider a homogeneous film on a Winkler foundation as a benchmark.
For the homogeneous case $B(x)\equiv B_0$ and a Winkler foundation with $K(q)\equiv K$, linear stability analysis gives the classical dispersion relation $D_0(q;N)=B_0 q^4-Nq^2+K=0$.
The corresponding critical load is $N_c=2\sqrt{B_0K}$, and the critical wavenumber is $q_c=(K/B_0)^{1/4}$, which defines the reference wrinkling wavelength $\bar{\lambda}=2\pi/q_c$ for the homogeneous system.
In the benchmark calculation,  we take $L=4\pi$, $K=1$, and $B_0=1$, which yields the theoretical predictions $N_c=2.0$ and $\bar{\lambda}=2\pi$.

\begin{figure}[ht]
    \centering
    \includegraphics[width=1.0\textwidth]{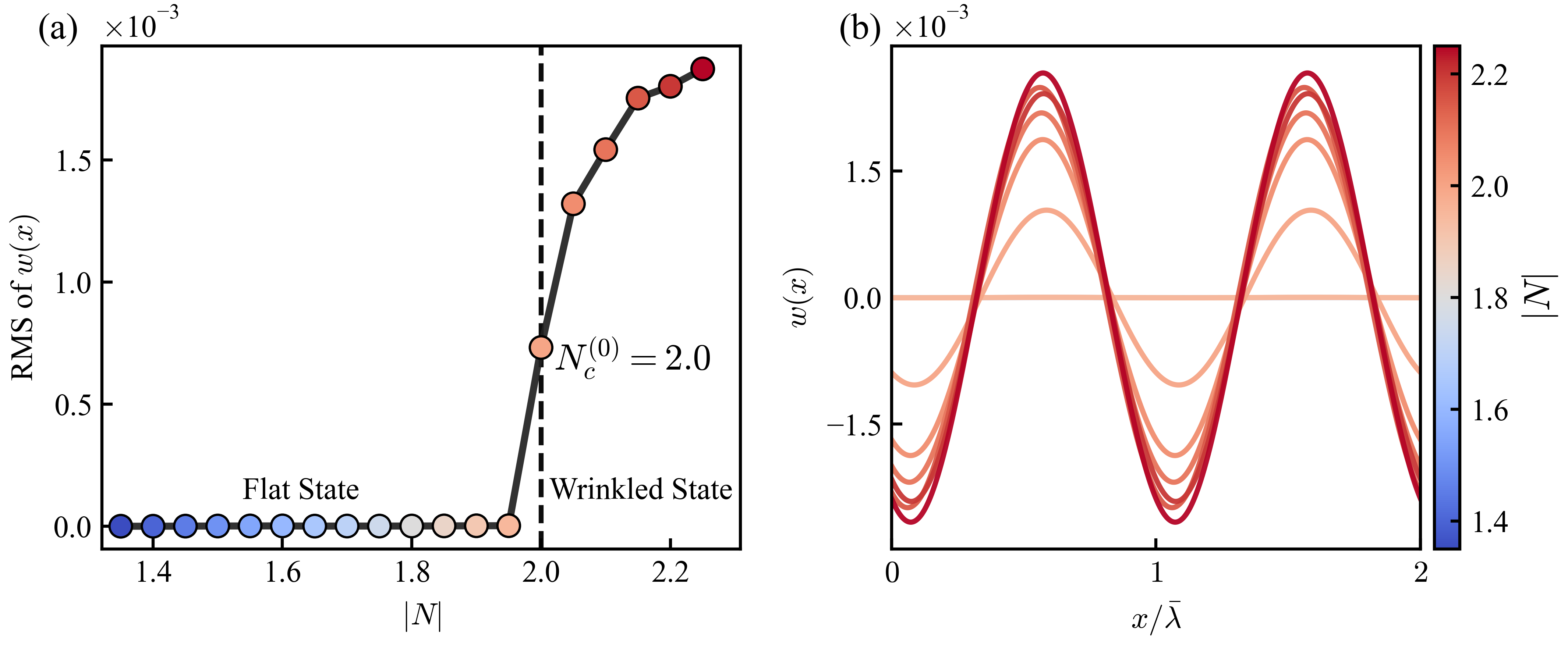}
    \caption{Homogeneous benchmark for the film--substrate system. 
    The film has a constant bending stiffness $B(x)\equiv B_0$, and the substrate stiffness is $K(q)\equiv K$.
    In this benchmark, $B_0=1$, $K=1$, and $L=4\pi$, giving the analytical critical load $N_c^{(0)}=2\sqrt{B_0K}=2.0$, the critical wavenumber $q_c=(K/B_0)^{1/4}=1.0$, and the reference wrinkling wavelength $\bar{\lambda}=2\pi/q_c=2\pi$.
    (a) Root-mean-square deflection as a function of the magnitude of the compressive load $|N|$.
    The dashed vertical line marks the analytical critical load $N_c^{(0)}=2.0$.
    Below this value, the film remains in the flat state with nearly zero deflection; above it, the RMS amplitude increases rapidly, indicating the onset of wrinkling.
    The marker colors represent the load magnitude and are consistent with the color scale used in panel (b).
    (b) Evolution of the out-of-plane displacement profiles $w(x)$ at different compressive loads.
    The horizontal coordinate is normalized by the reference wrinkling wavelength $\bar{\lambda}$, so that the domain $L=4\pi$ corresponds to two reference wavelengths.
    After the instability onset, the displacement develops a sinusoidal wrinkle pattern with wavelength consistent with $\bar{\lambda}$.
    }
    \label{fig:homogeneous_benchmark}
\end{figure}

Fig.~\ref{fig:homogeneous_benchmark}(a)shows the root-mean-square deflection as a function of the magnitude of the compressive load $|N|$. 
For $|N|<N_c^{(0)}$, the RMS amplitude remains nearly zero, indicating that the film stays in the flat state.
Once the load reaches the analytical threshold marked by the vertical dashed line, the RMS amplitude increases rapidly, signaling the onset of wrinkling instability.
The agreement between the numerical transition and $N_c^{(0)}=2.0$ confirms that the Fourier spectral implementation correctly captures the critical load of the homogeneous benchmark.
Fig.~\ref{fig:homogeneous_benchmark}(b) shows the corresponding out-of-plane displacement profiles at different load levels.
The horizontal coordinate is normalized by the reference wavelength $\bar{\lambda}$, so that the computational domain $L=4\pi$ spans two reference wavelengths.
Above the instability threshold, the film develops a nearly sinusoidal wrinkle mode with two periods over the domain.
This mode shape is consistent with the analytical prediction $q_c=1.0$ and $\bar{\lambda}=2\pi$.
Therefore, the homogeneous benchmark verifies both the critical load and the wavelength selection of the numerical method, providing a reference for the subsequent analysis of randomly heterogeneous systems.

In addition, we compute the critical point for the same benchmark using the generalized eigenvalue method. 
In this approach, the smallest positive eigenvalue gives the critical load and the associated eigenvector gives the critical mode. The result, $N_c=2.0$ and $q_c=1.0$, agrees exactly with the analytical solution. 
Therefore, the eigenvalue method provides a strict reference for the linear instability threshold, while the spectral continuation accurately reproduces the early post-buckling evolution.

\subsection{Effect of the Contrast Parameter $\varepsilon$ on the Critical Load}
\label{subsec:contrast_effect_critical_load}
This subsection investigates how the stiffness-contrast parameter affects the instability threshold. 
The analysis begins with the weak-contrast regime, where an asymptotic quadratic scaling can be derived, and then proceeds to finite and strong contrasts, where higher-order connected statistics become important. 
The SCE predictions are compared with generalized eigenvalue calculations to assess the accuracy of the second- and third-order truncations.

\subsubsection{Quadratic Scaling of the Critical Load in the Weak-Contrast Regime}
\label{subsec:weak_scaling}
We first consider the weak-contrast regime for the Winkler substrate with \(K(q)\equiv K\).
When $\varepsilon$ is small, the random stiffness fluctuation can be regarded as a perturbation to the homogeneous film.
The shift of the critical load is defined as
\begin{equation}
\Delta N_c(\varepsilon,\alpha)
=
N_c^{(0)}-N_c(\varepsilon,\alpha),
\label{eq:Delta_Nc_def_results}
\end{equation}
where $N_c^{(0)}$ is the critical load of the homogeneous reference system, and $N_c(\varepsilon,\alpha)$ is the critical load of the randomly heterogeneous system.

As derived in ~\ref{app:weak_scaling}, the effective susceptibility admits the weak-contrast expansion
\begin{equation}
L_e(q)
=
\varepsilon^2 \Lambda_2(q)
+
\varepsilon^4 \Lambda_4(q)
+
\mathcal O(\varepsilon^6).
\label{eq:Le_scaling_main}
\end{equation}

Substituting Eq.~\eqref{eq:Le_scaling_main} into the perturbation expansion of the effective dispersion relation gives
\begin{equation}
N(q)
=
N_0(q)
+
\varepsilon^2 N_2(q)
+
\varepsilon^4 N_4(q)
+
\mathcal O(\varepsilon^6),
\label{eq:N_expand_main}
\end{equation}
where
\begin{equation}
N_0(q)=B_0 q^2+\frac{K}{q^2},
\qquad
N_2(q)=q^2\Lambda_2(q),
\qquad
N_4(q)=q^2\Lambda_4(q)+\frac{q^2}{B_0}\Lambda_2(q)^2.
\end{equation}

A perturbation analysis around the homogeneous minimizer $q_0=(K/B_0)^{1/4}$ then gives
\begin{equation}
\Delta N_c
=
K_2(\alpha)\varepsilon^2
+
K_4(\alpha)\varepsilon^4
+
\mathcal O(\varepsilon^6),
\label{eq:deltaNc_expand}
\end{equation}
with
\begin{equation}
K_2(\alpha)
=
-q_0^2\Lambda_2(q_0),
\label{eq:K2_main}
\end{equation}
and
\begin{equation}
K_4(\alpha)
=
- q_0^2 \Lambda_4(q_0)
-\frac{q_0^2}{B_0}\Lambda_2(q_0)^2
+\frac{\left[2q_0\Lambda_2(q_0)+q_0^2\Lambda_2'(q_0)\right]^2}{16B_0}.
\label{eq:K4_main}
\end{equation}

Therefore, in the weak-contrast limit, the critical-load reduction follows the quadratic scaling
\begin{equation}
\Delta N_c\sim K_2(\alpha)\varepsilon^2,
\qquad
\varepsilon\to 0,
\label{eq:deltaNc_quad}
\end{equation}
The absence of a first-order correction reflects the zero-mean nature of the stiffness fluctuation, while the leading reduction is governed by second-order statistics.

The second-order and fourth-order approximations are defined as
\begin{equation}
\Delta N_c^{(2)}=K_2(\alpha)\varepsilon^2,
\end{equation}
and
\begin{equation}
\Delta N_c^{(4)}
=
K_2(\alpha)\varepsilon^2
+
K_4(\alpha)\varepsilon^4,
\end{equation}
respectively.
Their relative errors with respect to the full SCE result are evaluated by
\begin{equation}
\mathrm{Error}^{(m)}
=
\left|
\frac{\Delta N_c^{(m)}-\Delta N_c^{\mathrm{SCE}}}
{\Delta N_c^{\mathrm{SCE}}}
\right|\times 100\%,
\qquad m=2,4 .
\end{equation}

\begin{figure}[H]
    \centering
    \includegraphics[width=1.0\textwidth]{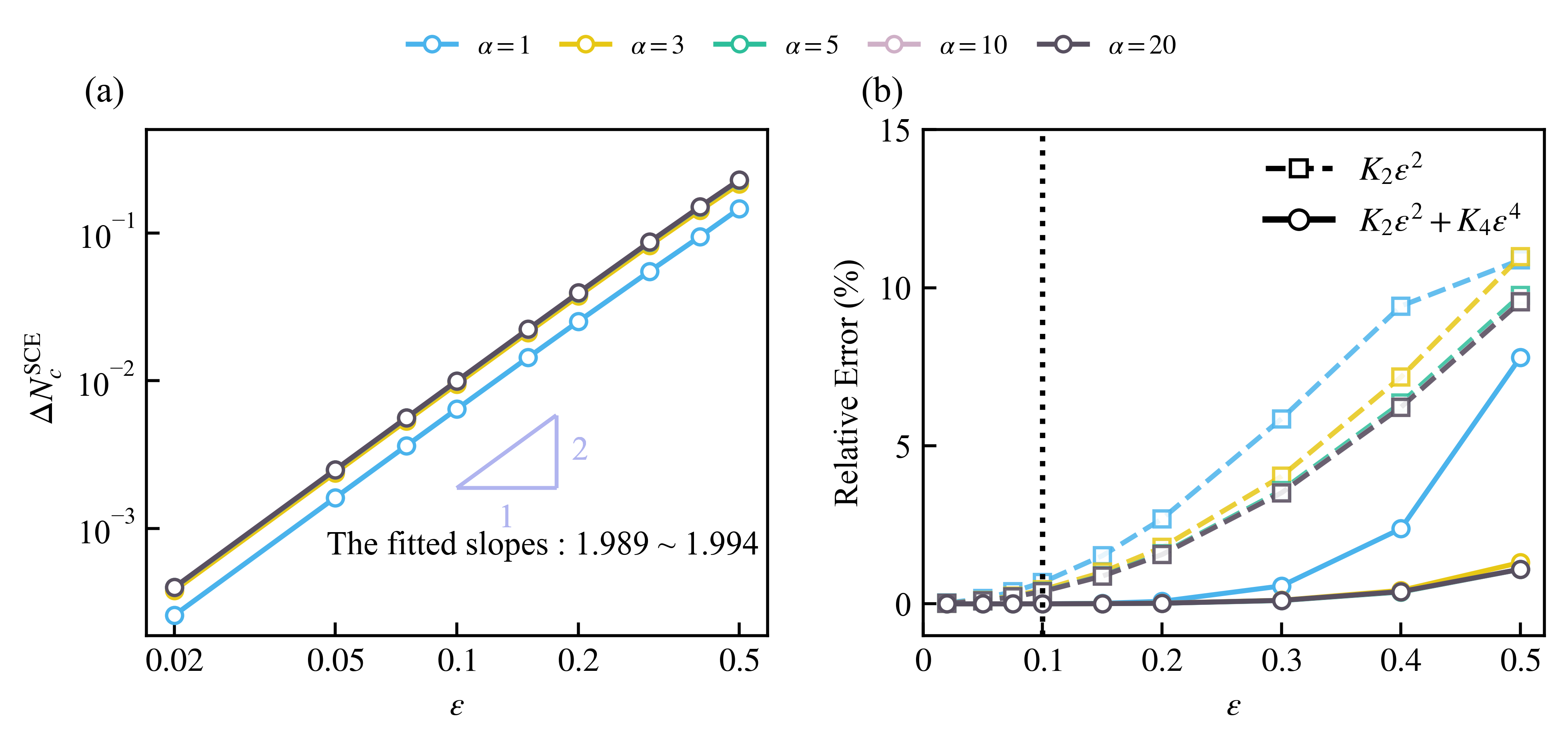}
    \caption{Weak-contrast scaling of the critical-load reduction predicted by the strong-contrast expansion for a Winkler substrate.
    (a)Log--log plot of the critical-load reduction  $\Delta N_c^{\mathrm{SCE}}=N_c^{(0)}-N_c$ as a function of the contrast parameter $\varepsilon$ for different spectral exponents $\alpha$.  
    The reference  triangle indicates the asymptotic scaling  $\Delta N_c\propto \varepsilon^2$ in the weak-contrast limit.
    (b) Relative errors of the second-order and fourth-order weak-contrast truncations, $\Delta N_c^{(2)}=K_2\varepsilon^2$ and  $\Delta N_c^{(4)}=K_2\varepsilon^2+K_4\varepsilon^4$, measured against the full SCE prediction.
    The vertical dashed line marks $\varepsilon=0.1$.
    For $\varepsilon\lesssim 0.1$,  both truncations remain accurate.
    The fourth-order correction extends the validity of the weak-contrast approximation to larger contrast levels.}
    \label{fig:weak_scaling}
\end{figure}

Fig.~\ref{fig:weak_scaling} examines the weak-contrast behavior of the critical-load reduction predicted by the SCE.
Fig.~\ref{fig:weak_scaling}(a) shows $\Delta N_c$ as a function of the contrast parameter $\varepsilon$ for different spectral exponents $\alpha$.
The results confirm the quadratic scaling for all spectral exponents considered.
All fitted slopes are close to $2$, in agreement with Eq.~\eqref{eq:deltaNc_quad}.
Fig.~\ref{fig:weak_scaling}(b) further quantifies the accuracy of the weak-contrast truncations.
It can be observed that for sufficiently small $\varepsilon$, especially for $\varepsilon\lesssim 0.1$, both second‑order and fourth‑order truncations provide good approximations to the full SCE results, with errors maintained at a low level. 
This confirms that the quadratic term dominates in the weak-contrast regime.
As $\varepsilon$ increases, the error of the second-order approximation gradually rises, implying that contributions from higher‑order terms become non‑negligible. 
By contrast, the fourth-order approximation remains accurate over a wider range of $\varepsilon$. However, when the contrast increases to moderate and strong levels, the weak‑contrast expansion is no longer adequate for accurately describing the critical load reduction, and the full strong‑contrast expansion is therefore required.

\subsubsection{Prediction of the Critical Load by the SCE Theory}
\label{subsubsec:strong_contrast_Nc}

In the finite- and strong-contrast regimes, the stiffness fluctuation becomes sufficiently large that the weak-contrast asymptotic expansion is no longer expected to provide a uniformly accurate description of the instability threshold. 
We therefore employ the strong-contrast expansion  and compare the second- and third-order SCE predictions with the critical loads obtained from the generalized eigenvalue calculations. 
This comparison assesses not only whether the SCE captures the reduction of the critical load, but also how the spectral exponent $\alpha$ affects the convergence of the theory.

\begin{figure}[H]
    \centering
    \includegraphics[width=1.0\textwidth]{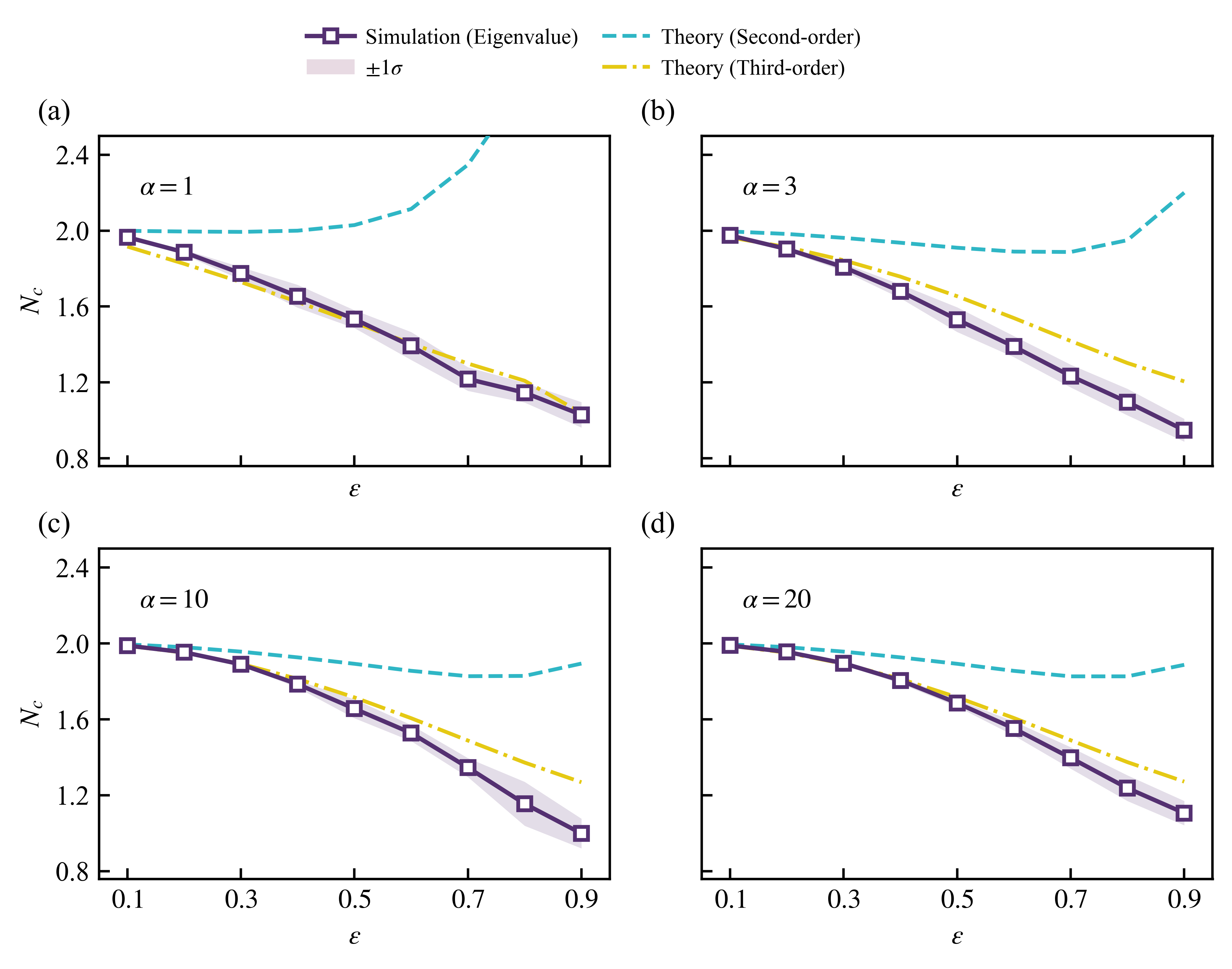}
    \caption{ Effect of stiffness contrast on the critical wrinkling load. 
    The critical load $N_c$ is plotted as a function of the contrast parameter $\varepsilon$ for different spectral exponents:
    (a) $\alpha=1$, (b) $\alpha=3$, (c) $\alpha=10$, and (d) $\alpha=20$. 
    The square points denote the mean critical loads obtained from generalized eigenvalue calculations, and the shaded bands indicate one standard deviation over different random realizations. 
    The dashed and dash-dotted curves denote the second- and third-order SCE predictions, respectively. Other parameters are fixed as $k_0=0.05$, $\sigma_k=2.0$, $B_m=1$, and $K=1$.}
    \label{fig:Nc_strong_contrast}
\end{figure}

Fig.~\ref{fig:Nc_strong_contrast} shows the dependence of the critical load $N_c$ on the contrast parameter $\varepsilon$ for four representative spectral exponents. 
For all values of $\alpha$, the eigenvalue results show a clear decrease in $N_c$ as $\varepsilon$ increases from the weak-contrast regime to the strong-contrast regime. 
Since the homogeneous reference value is $N_c^{(0)}=2$, this decrease indicates that random bending-stiffness heterogeneity lowers the onset load for wrinkling. 
Physically, a larger stiffness contrast produces more pronounced locally soft regions, which are more susceptible to bending deformation and therefore trigger instability earlier.

The comparison also shows the limitation of the second-order SCE at larger contrast.
At small $\varepsilon$, the second- and third-order predictions are close to each other and both remain near the eigenvalue results, which is consistent with the weak-contrast scaling discussed above.
As $\varepsilon$ increases, however, the second-order SCE increasingly overestimates the critical load and may even deviate from the monotonic decreasing trend observed in the numerical results. 
This behavior is especially evident when the contrast becomes strong, indicating that the second-order correction alone does not contain sufficient statistical information to represent the effect of strong stiffness heterogeneity.
By contrast, the third-order SCE gives a more accurate  prediction over the full range of $\varepsilon$. 
For $\alpha=1$, the third-order curve closely follows the eigenvalue results even at large contrast. 
For $\alpha=3$, $10$, and $20$, the third-order SCE still slightly overestimates $N_c$ in the high-contrast range, but it captures the overall decreasing trend and remains substantially closer to the simulations than the second-order theory. 
This improvement demonstrates that the third-order connected statistics included in the SCE are important for describing the instability threshold when the random stiffness field has moderate or strong contrast.

The dependence on $\alpha$ further indicates that the spectral structure of the random field influences the accuracy of the truncated expansion. 
Smaller $\alpha$ corresponds to a broader spectral distribution and stronger coupling among different spatial scales, so higher-order statistical contributions become more important. 
As $\alpha$ increases, the spectrum becomes more concentrated, and the discrepancy between the second- and third-order predictions is reduced. 
Overall, Fig.~\ref{fig:Nc_strong_contrast} confirms that stiffness heterogeneity systematically lowers the critical wrinkling load, and that the third-order SCE provides the more reliable theoretical approximation for strong random heterogeneity.

\subsection{Dominant Material Wavelength and Wrinkling Wavelength Selection}
\label{subsec:dominant_material_wavelength}
Random bending-stiffness heterogeneity affects not only the critical load, but also the wavelength selected at the onset of wrinkling.
This subsection examines wavelength selection in randomly heterogeneous films. 
In addition to the stiffness contrast, the dominant material wavelength introduces a second length scale that can interact with the natural wrinkling wavelength of the film--substrate system. 
The ratio between the dominant material wavelength and the harmonic-mean reference wavelength is used to characterize this interaction.

\subsubsection{Effects of Spectral Exponent $\alpha$ at Fixed Spectral Width }
We first examine the case in which the spectral width is fixed at $\sigma_k=2.0$.
Under this condition, changing the spectral exponent $\alpha$ modifies both the shape of the stiffness spectrum and the location of its dominant peak. 
Larger $\alpha$ shifts the spectral weight toward higher wavenumbers, corresponding to shorter material wavelengths and more rapidly varying stiffness fluctuations. 
This setting therefore provides a direct test of how stiffness contrast and spectral structure jointly affect the selected wrinkling wavelength.

\begin{figure}[H]
    \centering
    \includegraphics[width=1.0\textwidth]{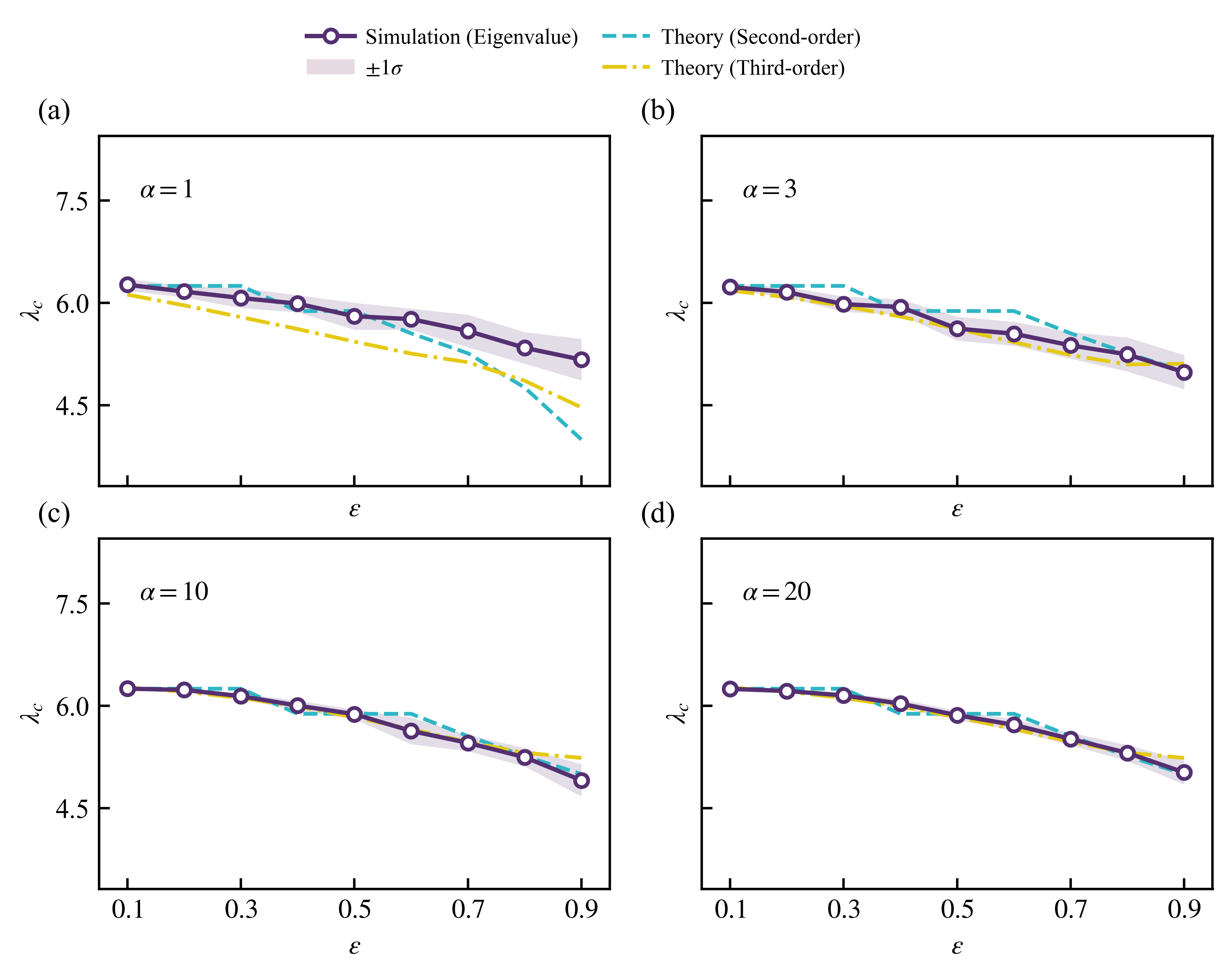}
    \caption{Effect of stiffness contrast on the actual wrinkling wavelength. 
    The critical wavelength $\lambda_c$ is plotted as a function of the contrast parameter $\varepsilon$ for different spectral exponents: (a) $\alpha=1$, (b) $\alpha=3$, (c) $\alpha=10$, and (d) $\alpha=20$. 
    Circle points denote the mean wavelengths obtained from generalized eigenvalue calculations, and the shaded bands indicate one standard deviation over different random realizations. 
    The dashed and dash-dotted curves denote the second- and third-order SCE predictions, respectively. Other parameters are fixed as $k_0=0.05$, $\sigma_k=2.0$, $B_m=1$, and $K=1$.}
    \label{fig:lambda_c_fixed_sigmak}
\end{figure}

Fig.~\ref{fig:lambda_c_fixed_sigmak} shows the actual wrinkling wavelength $\lambda_c$ as a function of  $\varepsilon$ at fixed $\sigma_k=2.0$ for different spectral exponents $\alpha$.
Overall, the results show that $\lambda_c$ decreases as $\varepsilon$ increases.
This indicates that random stiffness fluctuations not only reduce the critical load, but also shift the unstable mode toward higher wavenumbers. 
This trend can be understood from the role of locally soft regions.
A larger $\varepsilon$ amplifies low-stiffness regions, where out-of-plane bending deformation is easier to develop.
The SCE predictions capture this trend, with the third-order approximation generally giving a more stable prediction in the higher-contrast regime.

\subsubsection{Effects of Dominant Material Wavelength }

For the log-normal stiffness field generated by the exponential mapping Eq.~\eqref{eq:lognormal_B_main}, the harmonic mean of the bending stiffness is
\begin{equation}
B_H
=
\left\langle \frac{1}{B}\right\rangle^{-1}
=
B_m\exp(-\varepsilon^2),
\label{eq:BH_lognormal_results}
\end{equation}
and the corresponding harmonic-mean reference wavelength is
\begin{equation}
\lambda_H
=
2\pi\left(\frac{B_H}{K}\right)^{1/4}.
\label{eq:lambda_H_results}
\end{equation}

To separate the effect of the dominant material length from the change in spectral shape, the spectral width $\sigma_k$ is adjusted for each $\alpha$ so that the peak position $k^\ast$ of the stiffness power spectrum is prescribed.
For each spectral exponent $\alpha$, the spectral width $\sigma_k$ is adjusted according to Eq.~\eqref{eq:kstar_def}, so that different random stiffness fields share the same dominant material wavelength $\lambda^\ast=2\pi/k^\ast$. This construction keeps the leading material length scale fixed while allowing the spectral bandwidth and local fluctuation structure to vary with $\alpha$.

\begin{figure}[H]
    \centering
    \includegraphics[width=1.0\textwidth]{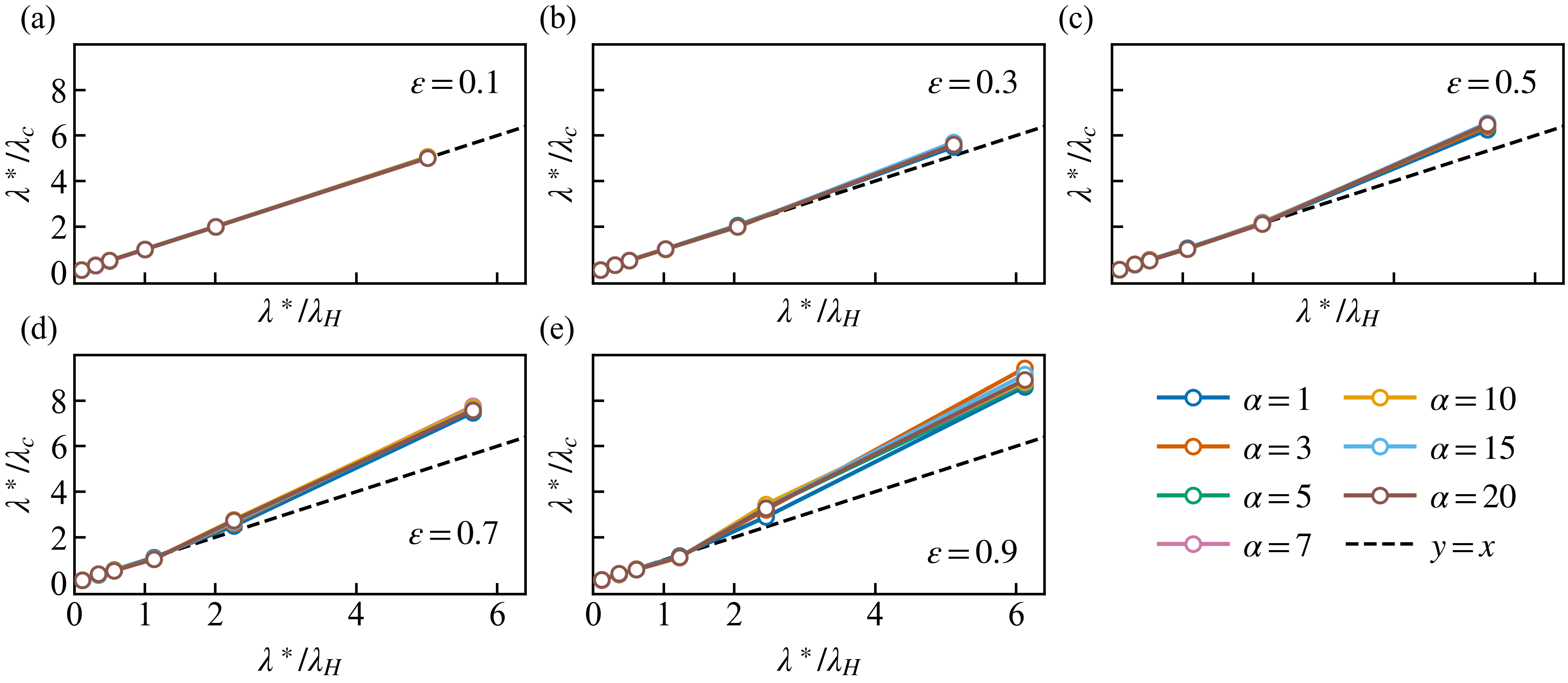}
    \caption{ Relation between the normalized dominant material wavelength $\lambda^\ast/\lambda_H$ and the actual wrinkling wavelength $\lambda^\ast/\lambda_c$, where $\lambda^\ast=2\pi/k^\ast$ is determined from the peak position of the stiffness power spectrum, $\lambda_H=2\pi(B_H/K)^{1/4}$ is the harmonic-mean reference wavelength, and $\lambda_c$ is the critical wrinkling wavelength obtained from the dominant wavenumber of the eigenmode. 
    Panels (a)--(e) correspond to increasing stiffness contrast: $\varepsilon=0.1$, $0.3$, $0.5$, $0.7$, and $0.9$, respectively. 
    Different colors denote different spectral exponents $\alpha$. For each $\alpha$, the spectral width $\sigma_k$ is adjusted to prescribe the same $k^\ast$, thereby separating the effect of the dominant material length scale from that of spectral bandwidth. 
    The dashed line denotes $y=x$, corresponding to $\lambda_c=\lambda_H$.
    Data above the dashed line indicate $\lambda_c<\lambda_H$. Other parameters are $k_0=0.05$, $B_m=1$, and $K=1$.
    }
    \label{fig:lambda_star_selection}
\end{figure}

Fig.~\ref{fig:lambda_star_selection} shows the relation between the normalized dominant material wavelength $\lambda^\ast/\lambda_H$ and the normalized wavelength ratio $\lambda^\ast/\lambda_c$ for different stiffness contrasts.
The dashed line $y=x$ corresponds to $\lambda_c=\lambda_H$. Therefore, data lying above this line indicate that the actual wrinkling wavelength is shorter than the harmonic-mean prediction, i.e., $\lambda_c<\lambda_H$.
In the weak-contrast regime, $\varepsilon=0.1$, the results remain close to the dashed line over the range of $\lambda^\ast/\lambda_H$ considered.
This indicates that, when stiffness fluctuations are weak, the harmonic-mean model provides a reliable estimate of the selected wrinkling wavelength.
As the contrast parameter $\varepsilon$ increases, the data progressively move above this line,  especially when $\lambda^\ast/\lambda_H$ approaches or exceeds unity.
This upward deviation means that the heterogeneous film selects a shorter wavelength than that predicted by the harmonic-mean reference model. 
Physically, strong stiffness contrast amplifies locally soft regions, which can accommodate bending deformation more easily and shift the unstable mode toward higher wavenumbers.

For  fixed $\varepsilon$ and a fixed $\lambda^\ast/\lambda_H$, results for different $\alpha$ collapse onto similar trends. 
This indicates that $\lambda^\ast/\lambda_H$ captures the primary length-scale effect of the random stiffness field, whereas $\alpha$ mainly affects spectral bandwidth and local fluctuation details.
When $\lambda^\ast/\lambda_H<1$,, stiffness fluctuations are averaged over one wrinkle period and the response remains close to the harmonic-mean prediction. 
When $\lambda^\ast$ becomes comparable to or larger than $\lambda_H$, the spatial organization of the stiffness field more directly interacts with the unstable mode, leading to stronger deviations from the harmonic-mean prediction.

\subsection{Mode Morphology and Localization}
\label{subsec:mode_localization}

For a homogeneous film, the critical mode is usually a nearly sinusoidal periodic pattern extended over the whole domain, with its Fourier energy concentrated around a single dominant wavenumber.
In contrast, the critical mode of a randomly heterogeneous film may exhibit spatial amplitude modulation, multi-wavenumber mixing, and localized wave-packet structures.
To quantify the spectral complexity of the critical mode, we introduce the normalized spectral entropy
\begin{equation}
H_{\rm spec}
=
-\frac{\sum_q p_q\ln p_q}{\ln M},
\qquad
p_q=
\frac{|\hat{w}(q)|^2}{\sum_q |\hat{w}(q)|^2},
\end{equation}
where $M$ is the number of nonzero frequency components included in the calculation.
When $H_{\rm spec}$ is close to zero, the Fourier energy is concentrated in only a few wavenumbers and the mode is close to a single-wavenumber sinusoidal pattern.
A larger value indicates stronger redistribution of Fourier energy among multiple wavenumber components, reflecting more pronounced mode mixing induced by material heterogeneity.

To quantify localization in physical space, we use the participation length
\begin{equation}
\ell_{\rm loc}
=
\frac{\left(\int |w(x)|^2\,dx\right)^2}
{\int |w(x)|^4\,dx},
\end{equation}
and define the normalized localization length as $P_{\rm loc}=\frac{\ell_{\rm loc}}{L}.$
Large $P_{\rm loc}$ indicates that the mode is distributed over a large spatial region, whereas  small $P_{\rm loc}$ indicates that the mode is localized within a small spatial region.

\begin{figure}[ht]
    \centering
    \includegraphics[width=1.0\textwidth]{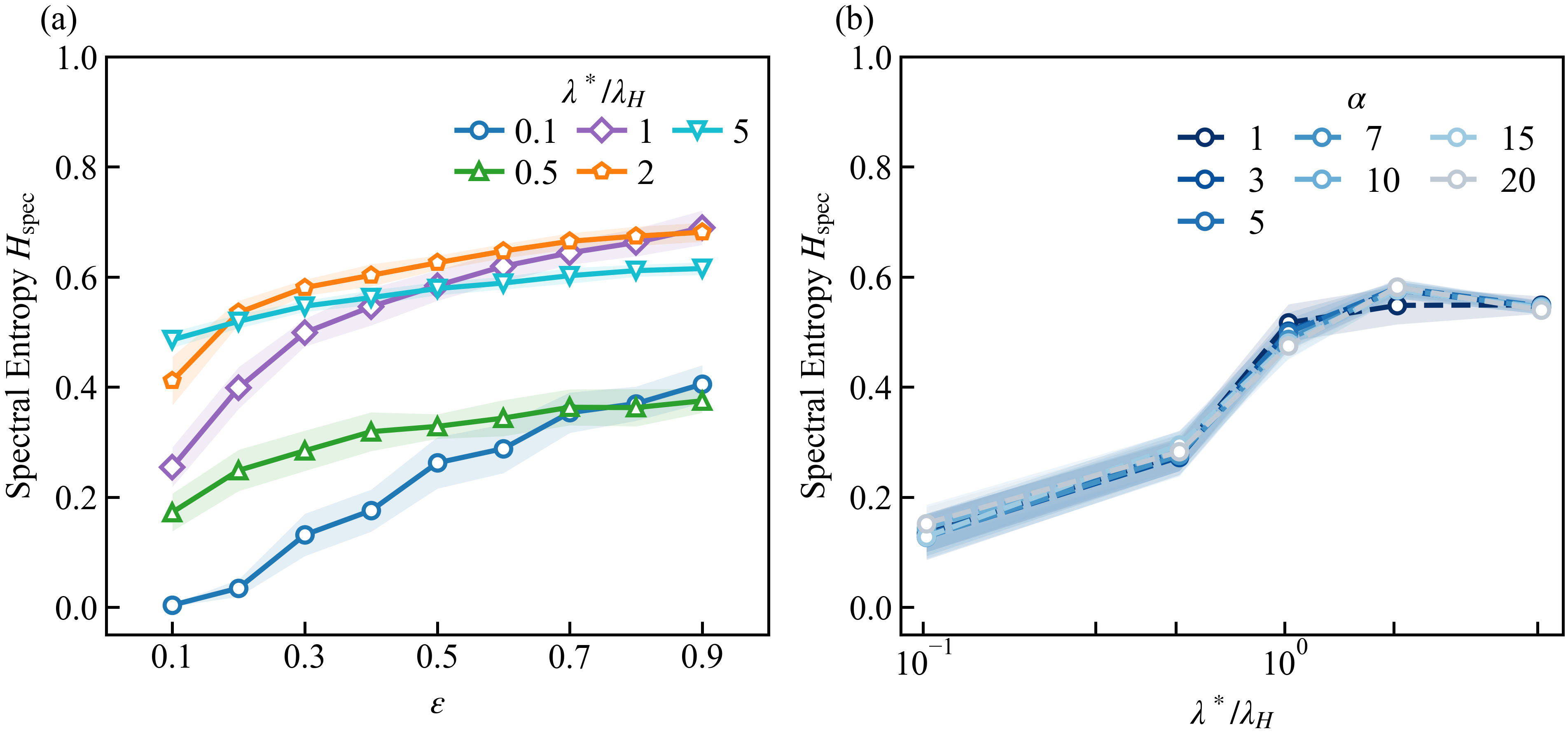}
    \caption{ Spectral complexity of critical wrinkling modes in randomly heterogeneous films. 
    The normalized spectral entropy $H_{\rm spec}$ measures the spreading of Fourier energy among different wavenumber components of the critical mode. 
    (a)  $H_{\rm spec}$ as a function of the stiffness-contrast parameter $\varepsilon$ for $\alpha=5$ and different nominal dominant material wavelengths $\lambda^\ast/\lambda_H$.
    Shaded bands denote $\pm 1$ standard deviation over random realizations. 
    (b) $H_{\rm spec}$ as a function of the harmonic-mean-normalized material wavelength $\lambda^\ast/\lambda_H$ at fixed contrast $\varepsilon=0.3$.
    Different curves correspond to different spectral exponents $\alpha$.
    }
    \label{fig:mode_entropy}
\end{figure}

Fig.~\ref{fig:mode_entropy} quantifies the spectral complexity of the critical wrinkling mode using the normalized spectral entropy $H_{\rm spec}$. 
Fig.~\ref{fig:mode_entropy}(a) shows that, for fixed spectral exponent $\alpha=5$, the spectral entropy generally increases with the contrast parameter $\varepsilon$. 
This trend indicates that stronger stiffness fluctuations cause the critical mode to depart progressively from the homogeneous single-wavenumber mode. 
For small $\lambda^\ast/\lambda_H$, such as $0.1$ or $0.5$, the spectral entropy remains relatively low.
This suggests that short-scale stiffness fluctuations are more effectively averaged over one wrinkle period, and the critical mode remains close to a single-wavenumber pattern.
By contrast, when the dominant material wavelength becomes comparable to or larger than the reference wrinkling wavelength, such as $\lambda^\ast/\lambda_H=1$, $2$, and $5$, the spectral entropy is much higher, indicating stronger coupling between the material length scale and the wrinkling mode and more pronounced multi-wavenumber mixing.
Fig.~\ref{fig:mode_entropy}(b) further examines the role of the spectral exponent $\alpha$ at fixed contrast $\varepsilon=0.3$. 
When the results are plotted against the harmonic-mean-normalized material wavelength $\lambda^\ast/\lambda_H$, the curves for different $\alpha$ cluster closely. 
This near collapse shows that the dominant material length scale, represented by $\lambda^\ast/\lambda_H$, controls the average spectral complexity more directly than the detailed spectral shape. 
The spectral exponent $\alpha$ mainly modifies the bandwidth and local fluctuation structure of the stiffness field, but its effect on the averaged spectral entropy is secondary once $\lambda^\ast/\lambda_H$ is prescribed.

\begin{figure}[H]
    \centering
    \includegraphics[width=1.0\textwidth]{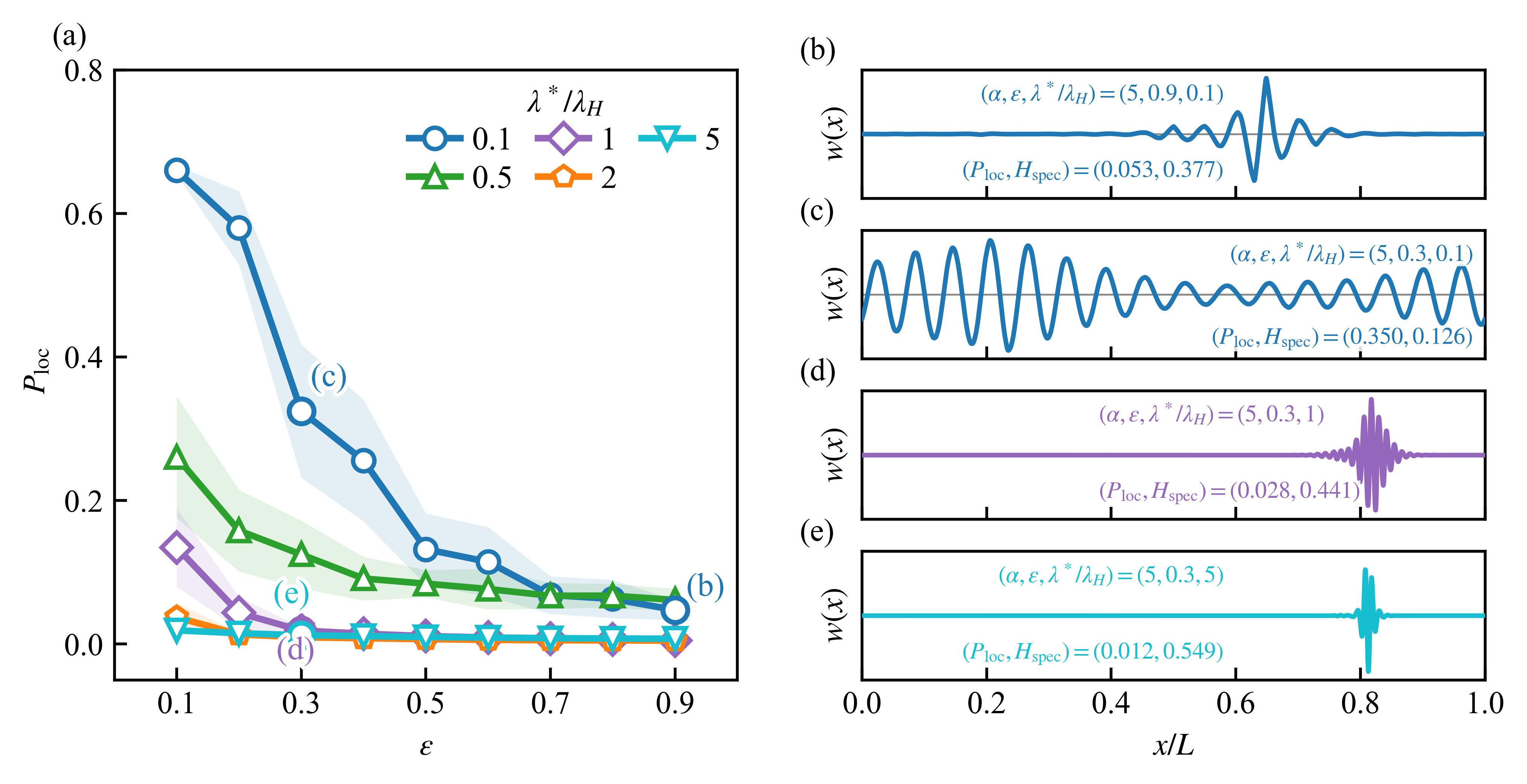}
    \caption{Spatial localization of critical wrinkling modes in randomly heterogeneous films. 
    (a) Normalized localization length $P_{\rm loc}$ as a function of $\varepsilon$ for different $\lambda^\ast/\lambda_H$ with $\alpha=5$.
    Smaller $P_{\rm loc}$ corresponds to stronger spatial concentration of the modal energy. 
    Shaded bands denote $\pm 1$ standard deviation over random realizations.
    (b)--(e)  Representative normalized critical modes $w(x)$ at the marked points in panel (a). 
    The values of $P_{\rm loc}$ and the normalized spectral entropy $H_{\rm spec}$ are reported in each panel.}
    \label{fig:mode_localization}
\end{figure}

Fig.~\ref{fig:mode_localization} further characterizes the spatial localization of the critical wrinkling mode using the normalized localization length $P_{\rm loc}$. 
A larger $P_{\rm loc}$ indicates that the mode extends over a substantial portion of the film, whereas a smaller value indicates that the modal energy is concentrated within a limited spatial region. 
Fig.~\ref{fig:mode_localization}(a) shows that $P_{\rm loc}$ decreases with increasing stiffness contrast $\varepsilon$ for all values of $\lambda^\ast/\lambda_H$, indicating stronger localization at higher stiffness contrast.
This trend is particularly clear for small $\lambda^\ast/\lambda_H$: at weak contrast, the mode may still extend over a relatively large portion of the domain, whereas at strong contrast it becomes concentrated in a localized region.
For larger $\lambda^\ast/\lambda_H$, $P_{\rm loc}$ is already small, indicating that stronger coupling between the material length scale and the mode scale favors localized wave-packet formation.

Figs.~\ref{fig:mode_localization}(b)--(e) show representative critical modes corresponding to the marked points in Fig.~\ref{fig:mode_localization}(a).
In Fig.~\ref{fig:mode_localization}(c), for $\varepsilon=0.3$ and $\lambda^\ast/\lambda_H=0.1$, the mode still shows an extended structure, with $P_{\rm loc}=0.350$ and a relatively low spectral entropy.
Increasing the contrast to $\varepsilon=0.9$ reduces $P_{\rm loc}$ to $0.053$, indicating that the deformation becomes concentrated in a localized wave packet. 
At the same contrast level $\varepsilon=0.3$, increasing $\lambda^\ast/\lambda_H$ to $1$ and $5$ further reduces $P_{\rm loc}$ to $0.028$ and $0.012$, respectively, while $H_{\rm spec}$ increases. 
Thus, random stiffness heterogeneity modifies the critical mode through two coupled mechanisms: it spreads modal energy over multiple wavenumber components and spatially traps the deformation around locally soft regions.

\subsection{Validation for Heterogeneous Films on Solid Substrate}
\label{subsec:solid_validation}

This subsection examines whether the scalar-kernel formulation remains applicable when the film is bonded to an elastic half-space substrate.
Unlike Winkler foundations, solid substrates exhibit nonlocal response and tangential--normal coupling at the interface. 
As discussed in Section~\ref{subsec:gov-equation}, this coupling is condensed into the effective scalar kernel $K_{\mathrm{eff}}(q)$, which can then be inserted into the same SCE-based dispersion criterion.

\begin{figure}[H]
    \centering
    \includegraphics[width=1.0\textwidth]{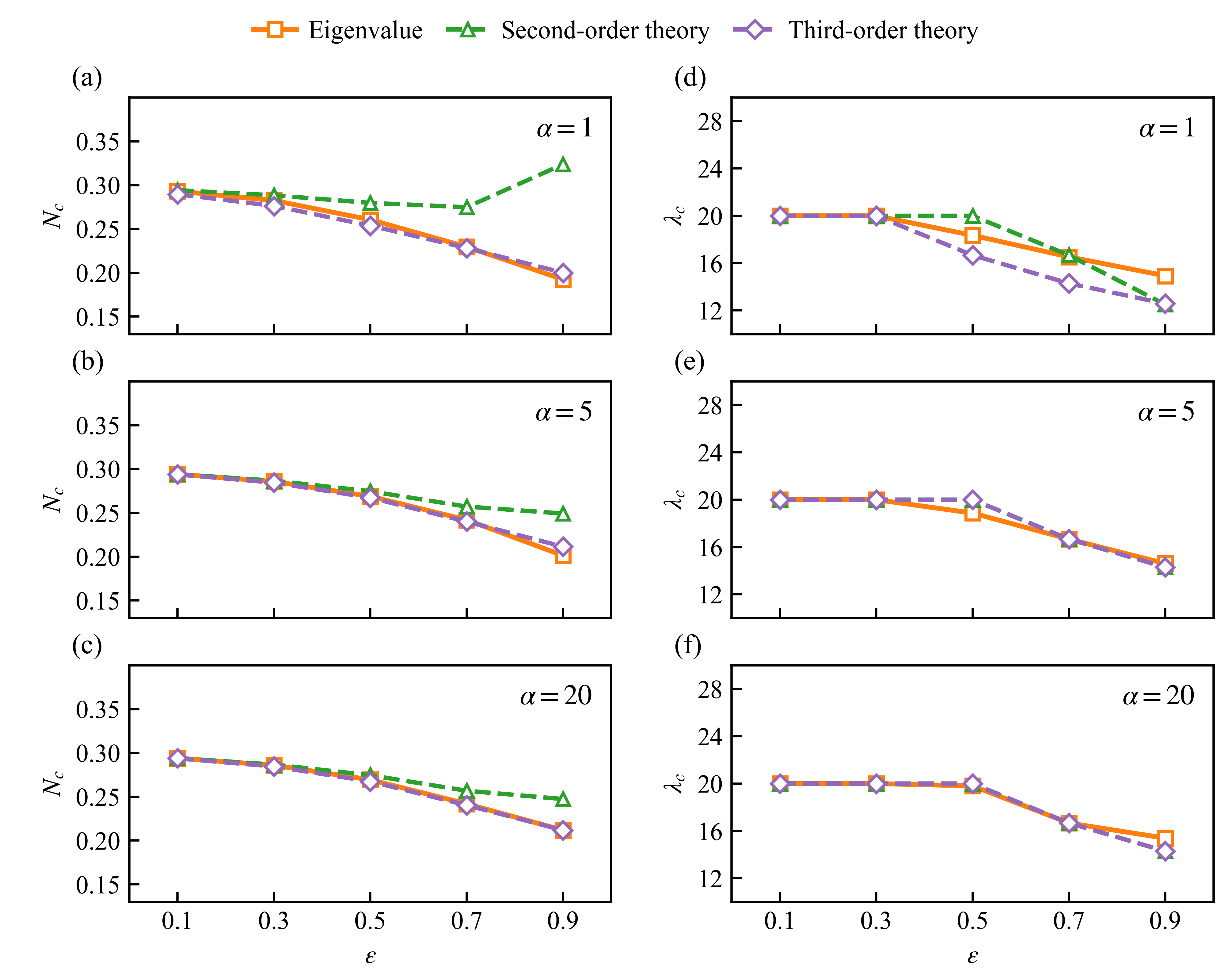}
    \caption{Numerical validation of the SCE framework for randomly heterogeneous films on an elastic half-space substrate. 
    Panels (a)--(c) show the critical load $N_c$ as a function of the contrast parameter $\varepsilon$, and panels (d)--(f) show the corresponding critical wrinkling wavelength $\lambda_c$. The spectral exponent is $\alpha=1$, $5$, and $20$ from top to bottom. 
   }
    \label{fig:solid_validation}
\end{figure}

Fig.~\ref{fig:solid_validation} compares the generalized eigenvalue results with the second- and third-order SCE predictions. 
The trends are consistent with those observed for Winkler substrates: increasing contrast lowers the critical load and decreases the selected wavelength. 
The third-order SCE gives substantially better agreement with the eigenvalue results than the second-order approximation, especially at larger contrast and smaller $\alpha$. 
The wavelength predictions also capture the main decreasing trend. 
These results validate both the scalar condensation procedure and the SCE framework for heterogeneous films on elastic half-space substrates.

\section{Conclusions}

This work developed a homogenized instability framework for one-dimensional film--substrate systems with randomly heterogeneous bending stiffness. 
The central contribution is to connect the statistical descriptors of a random stiffness field with the critical wrinkling load, selected wavelength, and morphology of the critical mode.
To this end, the heterogeneous stability problem was reformulated as a Lippmann--Schwinger equation for the curvature field. 
A local--nonlocal kernel decomposition and a cavity-field representation were then introduced to construct a strong-contrast expansion for the effective polarizability. 
This effective polarizability was incorporated into a Dyson-type dispersion relation, whose minimization gives the critical load and critical wavenumber. 
The formulation accommodates both Winkler or liquid foundations and elastic half-space substrates through a unified scalar substrate kernel.
For exponentially mapped Gaussian bending-stiffness fields, the two- and three-point connected statistics required by the SCE were obtained in closed form. 
This provides a calculable link between the covariance structure of the underlying Gaussian field, the contrast parameter of the log-normal stiffness distribution, and the instability response of the film.

The results show that random stiffness heterogeneity lowers the wrinkling threshold and shifts the unstable mode toward higher wavenumbers, resulting in shorter wrinkling wavelengths. 
In the weak-contrast regime, the critical-load reduction follows the quadratic scaling $\Delta N_c\sim\varepsilon^2$, indicating that the leading correction is governed by second-order statistics. 
At moderate and strong contrast, higher-order connected correlations become important, and the third-order SCE provides more robust predictions than the second-order approximation. 
The results also clarify the role of the ratio between the dominant material wavelength $\lambda^\ast$ and the harmonic-mean reference wavelength $\lambda_H$ in wavelength selection.
When $\lambda^\ast\ll\lambda_H$, stiffness fluctuations are effectively averaged within one wrinkle period and the harmonic-mean prediction remains accurate.
When $\lambda^\ast$ becomes comparable to or larger than $\lambda_H$, random heterogeneity promotes multi-wavenumber mixing and spatial localization of the critical mode, as reflected by increased spectral entropy and reduced participation length.

The comparisons with generalized eigenvalue calculations and Fourier spectral simulations demonstrate that the proposed framework captures the main instability trends over a broad range of stiffness contrasts and material length scales. 
The extension to elastic half-space substrates further shows that the scalar-kernel condensation preserves the predictive capability of the SCE for nonlocal substrate responses at the linear-instability level. 
It should be emphasized, however, that the present study is theoretical and numerical in scope.
Within this scope, this work provides a tractable homogenization theory for predicting how manufacturing-induced randomness modifies wrinkling instability in statistically heterogeneous film--substrate systems. 
In particular, the framework quantifies the changes in critical load and selected wavelength caused by stiffness fluctuations, which are key indicators for reliability assessment and structural design of heterogeneous thin-film systems.  
Future work may  include the two-dimensional wrinkling patterns, post-buckling evolution, non-Gaussian stiffness fields, and fully coupled matrix-valued strong-contrast formulations, together with experimental validation.


\section*{CRediT authorship contribution statement}
\textbf{Xinyu Xing}:  Writing – original draft, Visualization, Conceptualization, Methodology, Investigation.  
\textbf{Liyu Zhong}:  Writing – review and editing, Methodology, Investigation, Validation.
\textbf{Feng Deng}: Formal analysis; Software; Investigation; Writing – review and editing. 
\textbf{Sheng Mao}: Conceptualization, Supervision, Funding acquisition, Writing – review and editing.

\section*{Declaration of competing interest}
The authors declare that they have no known competing financial interests or personal relationships that could have appeared to influence the work reported in this paper.

\section*{Acknowledgments}

This work was supported by the National Key Research and Development Program of China (Grant No. 2023YFA1008900) and the National Natural Science Foundation of China (Grant  Nos. 12272005 and 12561160123). 
The authors also acknowledge the support from the high-performance computing platform of Peking University.

\appendix

\section{Condensation and Equivalent Normal Stiffness $K_{\mathrm{eff}}(q)$}
\label{app:Keff}
This appendix derives the scalar substrate kernel used for elastic half-space substrates. 
The purpose is to eliminate the in-plane displacement at the linear-instability level and obtain an equivalent normal stiffness \(K_{\mathrm{eff}}(q)\). 
This reduction allows the nonlocal solid-substrate response to be incorporated into the same scalar dispersion framework used for Winkler and liquid foundations.

Under linearized conditions, the membrane force in the film is approximated by
\begin{equation}
N \simeq C_0 u',
\end{equation}
where $C_0$ denotes the reference in-plane membrane stiffness of the film. 
For a homogeneous film,
\begin{equation}
C_0=\bar E_0 h, 
\qquad 
\bar E_0=\frac{E_0}{1-\nu^2}.
\end{equation}
For a non-uniform film, $C_0$ is taken as an appropriate effective or reference value.

In Fourier space, the in-plane equilibrium equation $N'=T_1$ gives
\begin{equation}
\hat T_1 = -\,C_0 q^2 \hat u.
\label{eq:T1_membrane_ch2}
\end{equation}

Combining Eq.~\eqref{eq:T1_membrane_ch2} with the first row of Eq.~\eqref{eq:D_matrix_def}, the in-plane displacement can be expressed in terms of the out-of-plane displacement as
\begin{equation}
\hat u = -\,\frac{D_{12}(q)}{C_0 q^2 + D_{11}(q)}\,\hat w.
\label{eq:u_condense_ch2}
\end{equation}

Substituting Eq.~\eqref{eq:u_condense_ch2} into the second row of Eq.~\eqref{eq:D_matrix_def} yields
\begin{equation}
\hat T_2 = K_{\mathrm{eff}}(q)\,\hat w,
\qquad
K_{\mathrm{eff}}(q)
= D_{22}(q) - \frac{D_{21}(q)\,D_{12}(q)}{C_0 q^2 + D_{11}(q)} .
\label{eq:Keff_ch2}
\end{equation}

For the one-dimensional plane-strain half-space kernel adopted in this work, Eqs.~\eqref{eq:D11_solid}--\eqref{eq:D21_solid}, Eq.~\eqref{eq:Keff_ch2} becomes
\begin{equation}
K_{\mathrm{eff}}(q)
=4C_{\mathrm{sub}}(1-\nu_s)|q|
-\frac{4C_{\mathrm{sub}}^2(1-2\nu_s)^2 q^2}{C_0 q^2 + 4C_{\mathrm{sub}}(1-\nu_s)|q|}.
\label{eq:Keff_explicit_ch2}
\end{equation}

Therefore, the coupled tangential--normal response of the elastic half-space is reduced to an equivalent scalar normal stiffness acting on the deflection field.

\section{Lippmann--Schwinger Equation for Curvature}
\label{app:LS-kappa}
For notational simplicity, the derivation below is written for a Winkler substrate. 
For a general Fourier-diagonal substrate response, the same derivation applies by replacing 
\(K\) with \(K(q)\) in the reference dispersion relation.
The bending stiffness is decomposed into its mean and fluctuating parts as
\begin{equation}
B(x)=B_0+\Delta B(x),
\end{equation}
where $B_0$ is the spatial average and $\Delta B(x)$ is a zero-mean fluctuation field.
Substituting this decomposition into the governing equation and moving the heterogeneous contribution to the right-hand side yields
\begin{equation}
(B_0 w'')'' + N w'' + K w = -\big(\Delta B(x)\, w''\big)''.
\end{equation}

Defining the differential operator for the reference medium
\begin{equation}
\mathcal L_0 := B_0\,\partial_x^4 + N\,\partial_x^2 + K,
\end{equation}
and the equivalent source term
\begin{equation}
f(x):=\big(\Delta B(x)\,w''(x)\big)'',
\end{equation}
the heterogeneous problem can be rewritten as
\begin{equation}\label{eq:perturbed_eq}
\mathcal L_0 w(x)=-f(x).
\end{equation}

Eq.~\eqref{eq:perturbed_eq} describes a uniform reference medium driven by a source induced by stiffness fluctuations. 
Its solution can be expressed in terms of the Green's function $G_0$ associated with $\mathcal L_0$, namely
\begin{equation}
\label{eq:w-conv}
w(x) = - \int G_0(x - \xi) f(\xi)  d\xi = - G_0 \convo f = - G_0 \convo \big( \Delta B \cdot w'' \big)'',
\end{equation}
where, $G_0(x)$ satisfies $\mathcal{L}_0 G_0(x) = \delta(x)$ with $\delta(x)$ denoting the Dirac delta function. 
In Fourier space,
\begin{equation}
\hat{G}_0(q) = \frac{1}{B_0 q^4 - N q^2 + K}.
\end{equation}

Introducing the curvature $\kappa(x)= w''(x)$, and applying the operator $\partial_x^2$ to Eq.~\eqref{eq:w-conv}, we obtain
\begin{align} 
\kappa(x)=-\,G_0*\big(\Delta B\,\kappa\big)^{(4)}(x).
\label{eq:kappa-intermediate}
\end{align}

By transferring the two derivatives from $\big(\Delta B\,\kappa\big)''$ to the Green's kernel, this relation can be written as
\begin{equation}
\kappa(x)
=
- \int \Xi^0(x-x')\,\big(\Delta B\,\kappa\big)(x')\,dx',
\end{equation}
where
\begin{equation}
\Xi^0(x-x')=\partial_x^2 \partial_{x'}^2 G_0(x-x').
\end{equation}

Accordingly,
\begin{equation}
\kappa = -\,\Xi^0 * \big(\Delta B\,\kappa\big).
\end{equation}
In Fourier space, this becomes
\begin{equation}
\hat{\kappa}(q)=-\hat{\Xi}^0(q)\,\widehat{\Delta B\,\kappa}(q),
\end{equation}
with
\begin{equation}
\hat{\Xi}^0(q)=q^4 \hat{G}_0(q).
\end{equation}

Therefore, the curvature satisfies the following Lippmann--Schwinger-type integral equation:
\begin{equation}
\kappa = -\,\Xi^0 * \tau,
\label{eq:kappa-LS}
\end{equation}
where
\begin{equation}
\tau=\Delta B\,\kappa
\end{equation}
is the polarization field associated with the stiffness heterogeneity.

\section{Kernel Decomposition, Cavity Field Definition and Local Fractional Relation}
\label{app:cavity}
This appendix explains the local--nonlocal decomposition of the curvature Green kernel and the resulting cavity-field definition. 
The singular local part of the kernel is separated from the regular nonlocal part, leading to a bounded local relation between the polarization and the cavity curvature field. 

For Eq.~\eqref{eq:kappa-LS}, note that the kernel $\widehat{\Xi^0}(q)$ satisfies
\begin{equation}
    \label{eq:Xi-lim}
\lim_{|q|\to\infty}\widehat{\Xi^0}(q)
=
\lim_{|q|\to\infty}\frac{q^4}{B_0 q^4 - N q^2 + K}
=
\frac{1}{B_0}.
\end{equation}
Since a constant in Fourier space corresponds to a Dirac delta term in real space, $\Xi^0$ can be naturally decomposed as
\begin{equation}\label{eq:Xi-split-Fourier}
\widehat{\Xi^0}(q)=D^0+\widehat{H^0}(q),
\end{equation}
where
\begin{equation}
D^0=\frac{1}{B_0},
\qquad
\widehat{H^0}(q)=\frac{q^4}{B_0 q^4-N q^2+K}-\frac{1}{B_0}.
\end{equation}
This decomposition separates the singular local contribution ($\delta$ -term) from the regular nonlocal part ($H^0$ term), which is the key step in constructing a bounded strong-contrast expansion.

Taking the inverse Fourier transform gives
\begin{equation}
\label{eq:Xi-split}
\Xi^0(x) \;=\; D^0\,\delta(x) \;+\; H^0(x).
\end{equation}

Substituting Eq.~\eqref{eq:Xi-split} into the Lippmann--Schwinger equation~\eqref{eq:kappa-LS} yields
\begin{equation}
\kappa=-(D^0\delta+H^0)*\tau=-D^0\,\tau-H^0*\tau.
\end{equation}
Accordingly, the cavity field associated with the curvature is defined as
\begin{equation}
\mathcal F_\kappa=-\,H^0 * \tau,
\end{equation}
so that
\begin{equation}
\label{eq:kappa-Fcav}
\kappa=\mathcal F_\kappa-D^0\,\tau.
\end{equation}

Eliminating $\kappa$ from $\tau=\Delta B\,\kappa$ and Eq.~\eqref{eq:kappa-Fcav} then gives the local fractional relation
\begin{equation}
\tau=\Delta B\,\kappa
=\frac{\Delta B}{1+D^0\Delta B}\,\mathcal F_\kappa.
\end{equation}
Then, we define the local susceptibility as 
\begin{equation}
    \label{eq:LB-def}
L_B(x)\;=\;\frac{\Delta B(x)}{1 + D^0\,\Delta B(x)},
\end{equation}
so that
\begin{equation}
\label{eq:LB-def2}
\tau(x) = L_B(x)\mathcal F_\kappa(x).
\end{equation}

\section{Derivation of the Third-Order Effective Polarizability $L_e$ via Connected Clusters}
\label{app:Le3-proof}

This appendix establishes Eq.~\eqref{eq:Le_3rd_main} in the main text. 
Under truncation at $\mathcal O(\mathcal H^2)$, the effective polarizability admits the expansion
\begin{equation}
L_e \;=\; a\,I \;+\; \Delta_2 \;+\; \Delta_3 \;+\; \mathcal O(\mathcal H^3),
\qquad a:=\langle L\rangle,\;\; L=a+\delta L,\;\;\langle \delta L\rangle=0.
\label{eq:Le_order3_appendix_goal}
\end{equation}
where $\Delta_2$ and $\Delta_3$ denote the connected contributions of second and third order in the sense of linked clusters, respectively.

The derivation proceeds in two steps. 
First, the averaged response operator $\langle S\rangle$ is reorganized into connected components. 
Second, a Neumann expansion is performed for
\begin{equation}
Q=\langle S\rangle^{-1}+\mathcal H,
\end{equation}
after which the result is substituted into $L_e=Q^{-1}$.

\subsection{Definitions of $S,\;Q,\;L_e$ }
From the cavity-field equation, Eq.~\eqref{eq:cavity_eq_main}, together with the local relation, Eq.~\eqref{eq:LB_main}, we obtain
\begin{equation}
\tau = L\,\mathcal F_{\rm ext} + L\,\mathcal H\,\tau,
\label{eq:tau_eq_app}
\end{equation}
where $L$  is the multiplicative operator defined by $L(x)=L_B(x)$, and $\mathcal H$ is the convolution operator.

The formal solution of Eq.~\eqref{eq:tau_eq_app} is
\begin{equation}
\tau = L\bigl[I-\mathcal H L\bigr]^{-1}\mathcal F_{\rm ext}
=S\,\mathcal F_{\rm ext},
\qquad
S = L + L\mathcal H L + L\mathcal H L\mathcal H L + \cdots.
\label{eq:S_def_app}
\end{equation}
Taking the ensemble average of Eq.~\eqref{eq:S_def_app} gives 
\begin{equation}
\langle\tau\rangle=\langle S\rangle\,\mathcal F_{\rm ext}.
\end{equation}
Upon ensemble averaging, the cavity field satisfies
\begin{equation}
\langle \mathcal F_\kappa \rangle
= \mathcal F_{\rm ext} + \mathcal H\,\langle \tau \rangle
= \bigl[\langle S\rangle^{-1}+\mathcal H\bigr]\langle \tau \rangle
= Q\,\langle \tau \rangle.
\label{eq:Q_def_app}
\end{equation}
which defines the operator
\begin{equation}
Q=\langle S\rangle^{-1}+\mathcal H.
\end{equation}

The effective susceptibility operator is then defined by
\begin{equation}
L_e=Q^{-1},
\end{equation}
so that
\begin{equation}
\langle \tau\rangle = L_e\,\langle \mathcal F_\kappa\rangle .
\label{eq:Le_def_app}
\end{equation}

\subsection{Linked-Cluster Reorganization of  $\langle S\rangle$ up to $\mathcal O(\mathcal H^2)$}
Truncating the series at $\mathcal O(\mathcal H^2)$ gives
\begin{equation}
S = L + L\mathcal H L + L\mathcal H L\mathcal H L + \mathcal O(\mathcal H^3).
\label{eq:S_trunc_app}
\end{equation}

Let $L=a+\delta L$ with $\langle\delta L\rangle=0$.
Under the linked-cluster (connected-cumulant) decomposition, all averages involving $\delta L$ are reorganized into connected contributions. 
The second- and third-order connected chains are defined as
\begin{align}
\Delta_2 &:= \bigl\langle \delta L(1)\,\mathcal H(1,2)\,\delta L(2)\bigr\rangle_{\rm conn},
\label{eq:Delta2-def-app}\\
\Delta_3 &:= \bigl\langle \delta L(1)\,\mathcal H(1,2)\,\delta L(2)\,
                       \mathcal H(2,3)\,\delta L(3)\bigr\rangle_{\rm conn}.
\label{eq:Delta3-def-app}
\end{align}
where $\langle\cdot\rangle_{\rm conn}$ denotes the connected part.

The linked-cluster reorganization of $\langle S\rangle$ up to $\mathcal O(\mathcal H^2)$ is therefore
\begin{equation}
\langle S\rangle
= a\,I
+ a^2\,\mathcal H
+ a^3\,\mathcal H^2
+ \Delta_2
+ a(\Delta_2\mathcal H+\mathcal H\Delta_2)
+ \Delta_3
+ \mathcal O(\mathcal H^3).
\label{eq:avgS_app}
\end{equation}
Here, the first three terms arise solely from the mean part $a$, whereas the remaining terms represent the connected fluctuation corrections through third order.

\subsection{Truncation of $Q$ and $L_e$}
From Eq.~\eqref{eq:Q_def_app}, we have 
\begin{equation}
Q=\langle S\rangle^{-1}+\mathcal H.
\end{equation}
Expanding the inverse of Eq.~\eqref{eq:avgS_app} in a Neumann series up to $\mathcal O(\mathcal H^2)$, and then substituting the result into $Q$, yields
\begin{equation}
Q
= \frac{1}{a}I-\frac{1}{a^2}(\Delta_2+\Delta_3)
+\frac{1}{a^3}\Delta_2^2
+ \mathcal O(\mathcal H^3).
\label{eq:Q_app2}
\end{equation}
Here, $\Delta_2=\mathcal O(\mathcal H)$ and $\Delta_3=\mathcal O(\mathcal H^2)$. Under the linked-cluster reorganization, disconnected products such as $\Delta_2^2$ arise at intermediate stages of the expansion but do not alter the final connected form of $L_e$ through $\mathcal O(\mathcal H^2)$.
Inverting Eq.~\eqref{eq:Q_app2} then yields
\begin{equation}
L_e = Q^{-1}
= a\,I+\Delta_2+\Delta_3+\mathcal O(\mathcal H^3),
\label{eq:Le_final_app2}
\end{equation}
which recovers Eq.~\eqref{eq:Le_3rd_main} in the main text.

\section{Closed-form Derivation of $W_2$ and $W_3$ under the Exponential Mapping}
\label{app:W2W3-proof}

In this appendix, we derive Eqs.~\eqref{eq:W2_closed_main}--\eqref{eq:W3_closed_main} in the main text.
Let $g(x)$  be a stationary Gaussian field with zero mean and unit variance, with covariance function
\begin{equation}
C(r)=\langle g(0)g(r)\rangle .
\end{equation}
For convenience we denote
\begin{equation}
C_{ij}:=C(x_i-x_j).
\end{equation}

\subsection{From the Mean-Preserving Log-Normal Mapping to  $\delta L(g)$}
We adopt the mean-preserving mapping in the main text:
\begin{equation}
B(x)=B_m\exp\!\Big(\varepsilon g(x)-\tfrac12\varepsilon^2\Big).
\end{equation}
Taking the reference stiffness $B_0=B_m$, the local fractional transformation gives
\begin{equation}
L(x)=B_0\Bigl(1-\frac{B_0}{B(x)}\Bigr)
= B_m\Bigl[1-\exp\!\bigl(-\varepsilon g(x)+\tfrac12\varepsilon^2\bigr)\Bigr].
\end{equation}

Define
\begin{equation}
\beta=e^{\varepsilon^2/2},
\qquad
Y(x)=\exp\!\Bigl(-\varepsilon g(x)-\tfrac12\varepsilon^2\Bigr),
\end{equation}
for which
\begin{equation}
\langle Y(x)\rangle = 1 .
\label{eq:def_Y}
\end{equation}

Using this notation we obtain
\begin{equation}
L(x)=B_m\bigl(1-\beta^2 Y(x)\bigr),
\qquad
a:=\langle L\rangle=B_m(1-\beta^2),
\end{equation}
and therefore
\begin{equation}
\delta L(x)=L(x)-a
=-B_m\beta^2\bigl(Y(x)-1\bigr).
\label{eq:deltaL_in_Y}
\end{equation}

\subsection{Hermite Expansion}

Let $H_k$ denote the Hermite polynomials defined by the generating function
\begin{equation}
\exp\!\Bigl(tz-\tfrac12 t^2\Bigr)
=\sum_{k=0}^{\infty}\frac{t^k}{k!}H_k(z).
\label{eq:Hermite_gen}
\end{equation}
 
Setting $t=-\varepsilon$ gives
\begin{equation}
Y(x)=\exp\!\Bigl(-\varepsilon g(x)-\tfrac12\varepsilon^2\Bigr)
=\sum_{k=0}^{\infty}\frac{(-\varepsilon)^k}{k!}H_k(g(x)).
\label{eq:Y_Hermite}
\end{equation}

Substituting \eqref{eq:Y_Hermite} into \eqref{eq:deltaL_in_Y} gives the Hermite expansion
\begin{equation}
\delta L(x)
= -B_m\beta^2\sum_{k=1}^{\infty}\frac{(-\varepsilon)^k}{k!}H_k(g(x))
= \sum_{k=1}^{\infty} b_k\,H_k(g(x)),
\qquad
b_k = B_m\beta^2\,\frac{(-1)^{k-1}\varepsilon^k}{k!}.
\label{eq:bk_closed}
\end{equation}

\subsection{Wick/Hermite Contraction Formulas}

For any zero-mean, unit-variance bivariate Gaussian vector $(G_1,G_2)$ with covariance $\langle\cdot\rangle E[G_1G_2]=\rho$, the Hermite contraction identity holds:
\begin{equation}
\langle H_k(G_1)H_\ell(G_2)\rangle=\delta_{k\ell}\,k!\,\rho^k.
\label{eq:Hermite_contract_2pt}
\end{equation}

In the stationary-field case, we set $G_1=g(0)$ and $G_2=g(r)$, giving $\rho=C(r)$.

For the three-point case, we use the generating-function identity for a trivariate Gaussian $(g_0,g_1,g_2)$:
\begin{equation}
\Big\langle\prod_{i=0}^2 \exp\!\bigl(t_i g_i-\tfrac12 t_i^2\bigr)\Big\rangle
=\exp\!\Bigl(t_0t_1C_{01}+t_0t_2C_{02}+t_1t_2C_{12}\Bigr),
\end{equation}
Comparing the coefficients of $t_0^k t_1^\ell t_2^m$ yields explicit expressions for
$\langle H_k(g_0)H_\ell(g_1)H_m(g_2)\rangle$.

\subsection{Two-Point Connected Quantity $W_2$}

Using \eqref{eq:bk_closed} and the contraction formula \eqref{eq:Hermite_contract_2pt},
\begin{align}
W_2(r)
&=\langle \delta L(0)\delta L(r)\rangle
=\sum_{k,\ell\ge1} b_k b_\ell \langle H_k(g(0))H_\ell(g(r))\rangle \nonumber\\
&=\sum_{k\ge1} b_k^2\,k!\,C(r)^k
=B_m^2\beta^4\sum_{k\ge1}\frac{(\varepsilon^2 C(r))^k}{k!}
=B_m^2\beta^4\Bigl(e^{\varepsilon^2 C(r)}-1\Bigr),
\end{align}
which coincides with Eq.~\eqref{eq:W2_closed_main} in the main text.

\subsection{Three-Point Connected Quantity $W_3$}

From \eqref{eq:deltaL_in_Y} and $\langle Y(x)\rangle=1$, and noting that cumulants of order $n\ge2$
are invariant under constant shifts, we obtain
\begin{equation}
W_3(r_1,r_2)
=\langle \delta L(0)\delta L(r_1)\delta L(r_2)\rangle_{\rm conn}
=(-B_m\beta^2)^3\,\kappa_3\!\bigl(Y_0,Y_1,Y_2\bigr),
\label{eq:W3_from_cumulant}
\end{equation}
where $Y_i=Y(x_i)$ and $\kappa_3$ denotes the third-order cumulant.

For any finite set of points $\{x_i\}$, the Gaussian exponential moment identity gives
\begin{equation}
\Big\langle \exp\Bigl(-\varepsilon\sum_i g(x_i)\Bigr)\Big\rangle
=\exp\Bigl(\tfrac12\varepsilon^2\,\mathrm{Var}\bigl(\sum_i g(x_i)\bigr)\Bigr).
\end{equation}

Combining this with the normalization term $-\tfrac12\varepsilon^2$ in \eqref{eq:def_Y} yields
\begin{equation}
\langle Y_iY_j\rangle = e^{\varepsilon^2 C_{ij}},
\qquad
\langle Y_0Y_1Y_2\rangle = e^{\varepsilon^2(C_{01}+C_{02}+C_{12})}.
\label{eq:Y_moments}
\end{equation}

Since $\langle Y_i\rangle=1$,the third-order cumulant reduces to
\begin{equation}
\kappa_3(Y_0,Y_1,Y_2)
=\langle Y_0Y_1Y_2\rangle
-\langle Y_0Y_1\rangle-\langle Y_0Y_2\rangle-\langle Y_1Y_2\rangle
+2.
\end{equation}

Substituting \eqref{eq:Y_moments} and setting $x_0=0$, $x_1=r_1$, $x_2=r_2$ (with $C_{12}=C(r_{12})$) gives
\begin{equation}
W_3(r_1,r_2)
= -\,B_m^3\beta^6\Bigl(
e^{\varepsilon^2\,[C(r_1)+C(r_2)+C(r_{12})]}
- e^{\varepsilon^2 C(r_1)}-e^{\varepsilon^2 C(r_2)}-e^{\varepsilon^2 C(r_{12})}
+2\Bigr),
\end{equation}
which is Eq.~\eqref{eq:W3_closed_main} in the main text.

This completes the closed-form derivation of the connected two-point and three-point quantities
$W_2$ and $W_3$ under the exponential mapping.

\section{Numerical Implementation Details}
\label{app:numerics}

This appendix summarizes the numerical implementation used to generate the random-field model introduced in Section~\ref{sec:random_W2W3}. 
The procedure includes the definition of discrete wavenumbers, spectral normalization, FFT-based sampling, and post-processing steps used to construct the Gaussian field $g(x)$ and the mapped field $B(x)$.

A periodic interval $[0,L)$ is discretized on a uniform grid with spacing $\Delta x$ and $N=L/\Delta x$ grid points,
\begin{equation}
x_j=j\Delta x, \qquad j=0,1,\dots,N-1 .
\end{equation}

The corresponding discrete wavenumbers are defined as
\begin{equation}
k_r=\frac{2\pi}{L}\left(r-\frac{N}{2}\right),
\qquad r=0,1,\dots,N-1 .
\end{equation}

The spatial structure of the random field is controlled by a prescribed power spectral density (PSD) $S_g(k)$. 
This work adopts the following form (see Eq.~\eqref{eq:psd_raw_main} in the main text):

\begin{equation}
S_{\alpha}^{\mathrm{raw}}(k)
=\left(|k|+k_0\right)^{\alpha}
\exp\!\left[-\frac{k^2}{2\sigma_k^2}\right],
\label{eq:psd_raw_1d}
\end{equation}
where $k_0>0$ prevents a singularity at $k=0$.

The spectrum is normalized in the discrete sense as
\begin{equation}
\mathcal{N}_{\alpha}=\sum_{r=0}^{N-1} S_{\alpha}^{\mathrm{raw}}(k_r),\qquad
S_g(k_r)=\frac{S_{\alpha}^{\mathrm{raw}}(k_r)}{\mathcal{N}_{\alpha}}.
\label{eq:psd_norm_1d}
\end{equation}
so that the discrete spectrum satisfies $\sum_r S_g(k_r)=1$.

The Gaussian random field is generated using a spectral‑filtering method. 
First, independent and identically distributed white noise $w(x_j)\sim\mathcal{N}(0,1)$ is generated in real space, and its discrete Fourier transform (DFT) $\hat w(k_r)$ is computed.
To ensure that the inverse transform produces a real-valued field, the spectral coefficients satisfy the conjugate symmetry condition
\begin{equation}
\hat g(-k)=\hat g^*(k).
\end{equation}

Filtering is then performed in the frequency domain using the square root of the target spectrum:
\begin{equation}
\hat{g}(k_r)=\sqrt{S_g(k_r)}\,\hat{w}(k_r),
\qquad
g(x_j)=\mathcal{F}^{-1}\!\left[\hat{g}(k_r)\right].
\label{eq:fourier_filtering}
\end{equation}

The resulting field is periodic over $[0,L)$. 
To reduce the mean and variance drift caused by finite sample length, each realization is standardized as
\begin{equation}
g(x_j)\leftarrow
\frac{g(x_j)-\langle g\rangle}{\sqrt{\mathrm{Var}(g)}},
\qquad
\langle g\rangle=\frac{1}{N}\sum_{j=0}^{N-1} g(x_j).
\end{equation}
so that the discrete realization satisfies $\langle g\rangle=0$ and $\mathrm{Var}(g)=1$.

After this standardization, the stiffness field is obtained through the mean-preserving exponential mapping
\begin{equation}
B(x_j)=B_m\exp\!\Bigl(\varepsilon g(x_j)-\tfrac12\varepsilon^2\Bigr),
\label{eq:B_discrete_mapping}
\end{equation}
which guarantees $\langle B(x_j)\rangle=B_m$ under the unit-variance Gaussian assumption for $g$.
Accordingly, the fluctuation amplitude is controlled by $\varepsilon$, while the spatial correlation structure of $B(x)$ is inherited from the underlying Gaussian field $g(x)$.

\section{Weak-contrast scaling of the instability threshold}
\label{app:weak_scaling}

In this appendix, we derive the scaling behavior of the instability threshold with respect to the contrast parameter $\varepsilon$ in the weak-contrast regime.

\subsection{Weak-contrast expansion of the effective susceptibility $L_e$}
\label{appG:Le_asymptotic}

\subsubsection{Expansion of the mean term $a$}
The effective susceptibility admits the strong-contrast expansion
(Eq.~\eqref{eq:Le_3rd_main})
\begin{equation}
L_e = a + \Delta_2 + \Delta_3 + \mathcal O(\mathcal H^3),
\end{equation}
where
\begin{equation}
a=\langle L\rangle = B_m(1-\beta^2),
\qquad
\beta=e^{\varepsilon^2/2}.
\end{equation}
Expanding $\beta$ for small $\varepsilon$ gives
\begin{equation}
\beta^2=e^{\varepsilon^2}
=1+\varepsilon^2+\frac12\varepsilon^4+\mathcal O(\varepsilon^6).
\end{equation}
Therefore
\begin{equation}
a
=
B_m(1-e^{\varepsilon^2})
=
- B_m \varepsilon^2
-\frac12 B_m \varepsilon^4
+\mathcal O(\varepsilon^6).
\end{equation}

\subsubsection{Expansion of the two-point contribution $\Delta_2$}
From the main text we have
\begin{equation}
W_2(r)=B_m^2\beta^4\left(e^{\varepsilon^2 C(r)}-1\right).
\end{equation}

We have
\begin{equation}
e^{\varepsilon^2 C(r)}-1
=
\varepsilon^2 C(r)
+\frac12\varepsilon^4 C(r)^2
+\mathcal O(\varepsilon^6),
\end{equation}
and 
\begin{equation}
\beta^4=e^{2 \varepsilon^2}
=1+2 \varepsilon^2+ 2\varepsilon^4+\mathcal O(\varepsilon^6).
\end{equation}

Multiplying these two terms together gives
\begin{equation}
W_2(r)
=
B_m^2   \left[  \varepsilon^2 C(r) + \varepsilon^4 \left( 2 C(r) +\frac12  C(r)^2   \right)   \right]
+\mathcal O(\varepsilon^6).
\end{equation}

In Fourier space,
\begin{equation}
\widehat W_2(q)=B_m^2 \left[ \varepsilon^2 \widehat C(q) + \varepsilon^4 \left( 2 \widehat C(q) +  \frac12 \widehat{C^2}(q) \right)\right]+\mathcal O(\varepsilon^6).
\end{equation}

Therefore
\begin{equation}
\Delta_2(q)
= \varepsilon^2 B_m^2 \widehat{{\mathcal H}}(q)\,\widehat{C}(q)
+ \varepsilon^4 B_m^2 \widehat{{\mathcal H}}(q)\left(2 \widehat{C}(q) + \frac{1}{2}\widehat{C^2}(q)\right)
+ \mathcal{O}(\varepsilon^6).
\end{equation}

\subsubsection{Expansion of the three-point contribution $\Delta_3$}
The three-point connected quantity is
\begin{equation}
W_3(r_1,r_2)
=
- B_m^3 \varepsilon^4
\left[
C(r_1)C(r_2)
+
C(r_1)C(r_{12})
+
C(r_2)C(r_{12})
\right]
+\mathcal O(\varepsilon^6),
\end{equation}
where $r_{12}=r_1-r_2$.
Let
\begin{equation}
G(r_1,r_2)=C(r_1)C(r_2)
+
C(r_1)C(r_{12})
+
C(r_2)C(r_{12}),
\end{equation}
then 
\begin{equation}
W_3(r_1,r_2)=-B_m^3\varepsilon^4G(r_1,r_2)+\mathcal O(\varepsilon^6).
\end{equation}

Performing a Fourier transform on $G$, under the variables $(p,q-p)$, we have 
\begin{equation}
\widehat{G}(p,q-p)
= \widehat{C}(p)\widehat{C}(q-p)
+ \widehat{C}(q)\widehat{C}(q-p)
+ \widehat{C}(p)\widehat{C}(q).
\end{equation}
so,
\begin{equation}
\widehat{W}_3(p,q-p)
= -B_m^3 \varepsilon^4
\left[
\widehat{C}(p)\widehat{C}(q-p)
+ \widehat{C}(q)\widehat{C}(q-p)
+ \widehat{C}(p)\widehat{C}(q)
\right]
+ \mathcal{O}(\varepsilon^6).
\end{equation}

Substitute into
\begin{equation}
\Delta_3(q)
=
\int_{\mathbb{R}} \frac{dp}{2\pi}\,
\widehat{{\mathcal H}}(p)\widehat{{\mathcal H}}(q-p)\,
\widehat{W}_3(p,q-p),
\end{equation}
we get
\begin{equation}
\Delta_3(q) = -B_m^3 \varepsilon^4 J(q) + \mathcal{O}(\varepsilon^6).
\end{equation}
where
\begin{equation}
J(q)
=
\int_{\mathbb R}\frac{dp}{2\pi}\,
\widehat{\mathcal H}(p)\widehat{\mathcal H}(q-p)
\Big[
\widehat C(p)\widehat C(q-p)
+\widehat C(q)\widehat C(q-p)
+\widehat C(p)\widehat C(q)
\Big].
\end{equation}

\subsubsection{Weak-contrast form of $L_e(q)$}
\label{appG:Le_form}

Combining the above expansions for $a$, $\Delta_2(q)$, and $\Delta_3(q)$, we obtain the weak-contrast form of the effective susceptibility
\begin{equation}
L_e(q)
=
a+\Delta_2(q)+\Delta_3(q)+\cdots.
\end{equation}

Therefore, for small $\varepsilon$

\begin{equation}
L_e(q)
=
\varepsilon^2 \Lambda_2(q)
+
\varepsilon^4 \Lambda_4(q)
+
\mathcal O(\varepsilon^6),
\label{eq:Le_scaling}
\end{equation}
where
\begin{equation}
\Lambda_2(q)
=
- B_m
+
B_m^2 \widehat{\mathcal H}(q)\widehat C(q),
\end{equation}
and
\begin{equation}
\Lambda_4(q)
=
-\frac12 B_m
+
B_m^2\widehat{\mathcal H}(q)
\left(2\widehat C(q)+\frac12\widehat{C^2}(q)\right)
-
B_m^3 J(q),
\end{equation}

\subsection{Perturbation of the dispersion relation}

From the effective closure condition (Eq.~\eqref{eq:closure_main})
the load parameter can be written as
\begin{equation}
N(q;N)
=
\frac{B_0 q^4 + K\,\phi(q;N)}
     {q^2\,\phi(q;N)},
\qquad
\phi(q;N) = 1-\frac{L_e(q;N)}{B_0}.
\label{eq:Nq_appendix}
\end{equation}

Since
\begin{equation}
\phi(q)^{-1}
=
\frac{1}{1-L_e(q)/B_0}
=
1+\frac{L_e(q)}{B_0}+\frac{L_e(q)^2}{B_0^2}+\mathcal{O}(\varepsilon^6),
\end{equation}
substituting this into the above expression gives
\begin{equation}
N(q)
=
B_0 q^2
\left(
1+\frac{L_e(q)}{B_0}+\frac{L_e(q)^2}{B_0^2}
\right)
+\frac{K}{q^2}
+\mathcal{O}(\varepsilon^6).
\label{eq:N_expansion_eps}
\end{equation}

Therefore,
\begin{equation}
N(q)=N_0(q)+\varepsilon^2 N_2(q)+\varepsilon^4 N_4(q)+\mathcal{O}(\varepsilon^6).
\end{equation}
where
\begin{empheq}[left=\empheqlbrace]{align}
N_0(q) &= B_0 q^2+\frac{K}{q^2}, \\
N_2(q) &= q^2\Lambda_2(q), \\
N_4(q) &= q^2\Lambda_4(q)+\frac{q^2}{B_0}\Lambda_2(q)^2.
\end{empheq}

Let the minimizer of the homogeneous system be $q_0$, satisfying
\begin{equation}
N_0'(q_0)=0.
\end{equation}

For
\begin{equation}
N_0(q)=B_0 q^2+\frac{K}{q^2},
\end{equation}
we have
\begin{equation}
q_0=\left(\frac{K}{B_0}\right)^{1/4},
\qquad
N_c^{(0)}=2\sqrt{B_0K},
\qquad
N_0''(q_0)=8B_0.
\end{equation}

We seek the perturbed critical wavenumber in the form
\begin{equation}
q_c=q_0+\varepsilon^2 q_1+\mathcal O(\varepsilon^4).
\end{equation}

From the stationarity condition $N'(q_c)=0$, one obtains
\begin{equation}
q_1=-\frac{N_2'(q_0)}{N_0''(q_0)}.
\end{equation}

Moreover,
\begin{equation}
N_2'(q)=2q\Lambda_2(q)+q^2\Lambda_2'(q),
\end{equation}
and therefore
\begin{equation}
q_1
=
-\frac{2q_0\Lambda_2(q_0)+q_0^2\Lambda_2'(q_0)}{8B_0}.
\end{equation}

Substituting $q_c=q_0+\varepsilon^2 q_1$ back into Eq.~\eqref{eq:N_expansion_eps} gives
\begin{equation}
N_c
=
N_c^{(0)}
+\varepsilon^2 N_2(q_0)
+\varepsilon^4
\left[
N_4(q_0)-\frac{N_2'(q_0)^2}{2N_0''(q_0)}
\right]
+\mathcal O(\varepsilon^6).
\end{equation}

Hence,
\begin{equation}
\Delta N_c
=
N_c^{(0)}-N_c
=
K_2(\alpha)\varepsilon^2
+
K_4(\alpha)\varepsilon^4
+
\mathcal O(\varepsilon^6),
\end{equation}
with
\begin{equation}
K_2(\alpha)=-N_2(q_0)=-q_0^2\Lambda_2(q_0),
\end{equation}
and
\begin{equation}
K_4(\alpha)
=
-N_4(q_0)
+
\frac{N_2'(q_0)^2}{2N_0''(q_0)}.
\end{equation}
Specifically,
\begin{equation}
K_4(\alpha)
=
- q_0^2 \Lambda_4(q_0)
-\frac{q_0^2}{B_0}\Lambda_2(q_0)^2
+\frac{\left[2q_0\Lambda_2(q_0)+q_0^2\Lambda_2'(q_0)\right]^2}{16B_0}.
\end{equation}

\bibliography{Wrinv2}

\end{document}